# A Learning Approach for Low-Complexity Optimization of Energy Efficiency in Multi-Carrier Wireless Networks

Salvatore D'Oro, *Member, IEEE*, Alessio Zappone, *Senior Member, IEEE*,
Sergio Palazzo, *Senior Member, IEEE*, Marco Lops, *Senior Member, IEEE*

***Abstract*—This paper proposes computationally efficient algorithms to maximize the energy efficiency in multi-carrier wireless interference networks, by a suitable allocation of the system radio resources, namely the transmit powers and subcarrier assignment. The problem is formulated as the maximization of the system global energy efficiency (GEE) subject to both maximum power and minimum rate constraints. This leads to a challenging non-convex fractional problem, which is tackled through an interplay of fractional programming, learning, and game theory. The proposed algorithmic framework is provably convergent and has a complexity linear in both the number of users and subcarriers, whereas other available solutions can only guarantee a polynomial complexity in the number of users and subcarriers. Numerical results show that the proposed method performs similarly as other, more complex, algorithms.**

## I. Introduction

The next generation of cellular networks will have to serve an unprecedented amount of wireless devices, which is forecast to reach 50 billions by 2020 [1]. This poses serious sustainable growth concerns, because in order to support these many devices, future networks will have to increase the supported data-rate by a factor 1000 as compared to current networks [2]. However, simply scaling up the transmit powers would lead to unmanageable energy demands and alarming levels of greenhouse gas emissions and electromagnetic pollution. Instead, it is widely accepted by the mobile community that the next generation of cellular networks will have to fulfill the 1000x data-rate requirement, while at the same time halving the energy consumption of today's networks [2]. This requires a 2000x increase of the bit-per-Joule energy efficiency.

A recent survey [3] identifies and discusses the more promising approaches to improve the energy efficiency of future 5G networks. Among others, one anticipated technique is to maximize the energy efficiency through a suitable allocation of the network radio resources. Otherwise stated, the traditional paradigm according to which communication networks are designed for data-rate maximization, should be rethought and the system radio resources must be allocated to maximize the global energy efficiency, defined as the ratio between the system throughput, and the total consumed power [4]. In the context of multi-carrier communications, the leading communication technology in current LTE networks, the radio resources to allocate are the transmit powers and subcarriers. First contributions in this direction have considered the simplifying assumption of orthogonal transmissions, which nulls out multi-user interference and leads to a problem that can be globally and efficiently solved by fractional programming theory [5]. However, exclusive subcarrier assignment does not appear a viable choice for future networks, as it is difficult to implement in multi-cell and heterogeneous scenarios, and because of the exponentially increasing spectrum demands [6]. Nevertheless, if multi-user interference is present, energy efficiency maximization becomes more complex, being in general an NP-hard problem. Hence, low-complexity solutions are required for practical applications.

One widely used approach to reduce complexity makes use of non-cooperative game theory [7]–[9]. In this context, instead of directly tackling the maximization of the system-wide energy efficiency with respect to all of the available network radio resources, the problem is formulated modeling the network nodes as rational agents that compete for individual energy efficiency maximization. Such an approach tackles the system-wide energy efficiency maximization problem by solving a set of user-dependent, convex or pseudo-convex problems, with a reduced set of optimization variables. This leads to a practical resource allocation algorithm, but typically suffers from a significant performance gap in terms of global network performance. In [10], energy-efficient power control in multi-carrier CDMA networks is studied, while in [11] the problem of non-cooperative power control in OFDMA networks is addressed. In [12] non-cooperative energy-efficient power control and receiver design is performed in relay-assisted networks.In [13], [14] the non-cooperative, energy-

S. Palazzo is with the CNIT Research Unit, Dipartimento di Ingegneria Elettrica, Elettronica e Informatica (DIEEI) at the University of Catania, Catania, Italy. At the time of this study, S. D'Oro was with the DIEEI at the University of Catania, Catania, Italy. A. Zappone is with the LANEAS group of the Laboratoire des Signaux et Systmes, CentraleSupélec, CNRS, Univ Paris Sud, Université Paris-Saclay, 3 rue Joliot Curie, Plateau du Moulon, 91192 Gif-sur-Yvette, France. (e-mail: alessio.zappone@l2s.centralesupelec.fr) M. Lops is with the Department of Electrical and Information Engineering of the University of Cassino and Southern Lazio, Cassino, Italy (e-mail: lops@unicas.it).

This work was supported in part by the European Commission through the H2020-MSCA IF-BESMART project under Grant Agreement 749336, and in part by the PRASG-GREEN5G project, funded by the University of Cassino and Southern Lazio.

This paper has been accepted for publication on *IEEE Transactions on Wireless Communications*. This is a preprint version of the accepted paper. The final paper will be available in short.

efficient power control problem is extended to account for minimum rate constraints. This problem is analyzed by employing the tool of generalized non-cooperative games [15]. In [13] the focus is on small-cell networks, whereas in [14] a more general setup is considered and a framework for non-cooperative energy efficiency maximization is provided, encompassing several 5G candidate technologies.

More recently, a more sophisticated, although still affordable, approach proposes to maximize the global energy efficiency of wireless interference networks, by merging fractional programming and sequential optimization theory [14], [16]. Again, the energy efficiency maximization problem is decomposed into a sequence of convex or pseudo-convex sub-problems. However, unlike game-theoretic approaches, the sub-problems considered in [14], [16] are not local problems aiming at maximizing the users' individual energy efficiency, but are instead network-wide problems in which all available radio resources are jointly optimized. This significantly improves the performance compared to game-theoretic approaches, but comes at the expense of a higher complexity, which becomes challenging in very large networks with many users and subcarriers.

Motivated by this scenario, the aim of this work is to develop a new centralized framework for energy efficiency optimization in wireless interference networks, which exhibits a comparable or lower complexity than available approaches, but near-optimal global energy efficiency performance. This is hereby achieved borrowing tools from machine learning [17]–[23]. Regarding this point, we stress that machine learning is typically used to develop online algorithms in dynamic environments in which the network nodes decide their resource allocation policy mainly based on past experience. This approach has been proved especially useful in fast-fading scenarios wherein a long-term performance measure is to be optimized. Contributions in this sense are [24], where a learning-based stochastic power control algorithm is proposed for non-cooperative energy efficiency in cognitive mesh networks, [25], that considers relay-based networks and employs a learning approach to determine mixed strategy Nash equilibrium (NE) points for the problem of non-cooperative energy-efficient power control. Energy-efficient learning-based power control is analyzed in [26] for femto-cell indoor scenarios. In [27] an exponential learning framework to maximize the system ergodic capacity in multi-carrier systems is proposed, whereas [28], [29] consider ergodic rate and sum energy efficiency maximization.

Nevertheless, while the main application of learning tools for radio resource allocation is in fast-fading environments, it was observed very recently in [30] that specific tools from learning theory can be successfully used also in slow-fading settings to reduce the complexity of the resource allocation process. Specifically, [30] shows that *stochastic learning with exponential mappings* can reduce the complexity of the rate optimization problem in a single-cell cognitive radio system. This work extends the approach in [30] considering energy efficiency optimization in multi-cell networks. Specifically, the following major contributions are made:

- A provably convergent power and subcarrier allocation algorithm for energy efficiency maximization is developed, merging tools from fractional programming, game theory, and learning theory. Similarly to available methods, the proposed algorithm operates by solving a sequence of convex sub-problems. However, we solve each sub-problem by an efficient fixed-point learning scheme, in which all update formulas are given in closed-form. This ensures a linear complexity in the number of subcarriers and users, whereas other available methods guarantee a polynomial complexity in the users and subcarriers number.
- Unlike many previous works, the developed optimization framework is able to handle not only maximum power constraints, but also minimum rate constraints, preserving all its salient properties. Moreover, despite the reduced complexity, an extensive numerical analysis shows that the proposed method suffers a negligible performance loss compared to more complex solutions based on sequential fractional programming theory.
- Unlike [30], the proposed method does not focus on only one specific communication system, but rather considers generic interference network topologies, with a more general signal to interference plus noise ratio (SINR) expression than that typically used in previous works. This allows modeling several relevant instances of communication systems, such as heterogeneous networks, multi-cell networks, massive MIMO systems with imperfect CSI and hardware-impairments, relay-assisted communications.

## II. System model and problem statement

Consider the uplink of an interference network with a set $\mathcal{K}$ of $K$ users communicating with $J$ receivers over $N$ subcarriers. User $k$'s SINR at its intended receiver, over subcarrier $n$, is:

$$\gamma_{k,n} = \frac{p_{k,n}\alpha_{k,n}}{\sigma_n^2 + \xi_{k,n}p_{k,n} + \sum_{\ell \neq k} p_{\ell,n}\beta_{\ell,n}^{(k)}} \ , \qquad (1)$$

with $p_{k,n}$ and $\sigma_n^2$ the $k$-th user's transmit power and the receive noise power over subcarrier $n$; $\{\alpha_{k,n}, \xi_{k,n}\}_{k,n}$, $\{\beta_{\ell,n}^{(k)}\}_{k,\ell,n}$ coefficients fulfilling the following two assumptions, for any $k$, $\ell$, $n$:

- they are non-negative real numbers which depend on global system parameters and channel gains, but not on the users' transmit powers.
- $\alpha_{k,n}$ and $\xi_{k,n}$ only depend on user $k$'s channel over resource block $n$, and possibly on system global parameters, while $\{\beta_{\ell,n}^{(k)}\}_{\ell}$ depend on the channel from transmitter $\ell$ to receiver $k$, over resource block $n$, and possibly on system global parameters.

The particular expressions of $\{\alpha_{k,n}\}_{k,n}$, $\{\xi_{k,n}\}_{k,n}$, and $\{\beta_{\ell,n}^{(k)}\}_{k,\ell,n}$ depend on the specific network under analysis, and we hasten to stress that many relevant instances of communication systems are modeled by (1), by suitably specifying $\{\alpha_{k,n}\}_{k,n}$, $\{\xi_{k,n}\}_{k,n}$, and $\{\beta_{\ell,n}^{(k)}\}_{k,\ell,n}$. Besides the simpler case in which $\xi_{k,n} = 0$ for all $k, n$, leading to the familiar SINR expression encountered in wireless communication systems, the presence of non-zero $\{\xi_{k,n}\}_{k,n}$ allows modeling several 5G candidate technologies. Examples in this sense include: practical massive MIMO networks subject to hardware impairments and/or imperfect channel estimation at the receiver [14]; relay-assisted networks [12], [31]; device-to-device (D2D) communications [32]. A detailed description of these case-studies is reported in [14]. In addition, other scenarios leading to the SINR in (1) are systems affected by inter-symbol interference and/or frequency-selective fading [33], [34].

The network GEE is the ratio of the achievable sum-rate over the total consumed power [4]:

$$\text{GEE} = \frac{W\sum_{k=1}^{K}\sum_{n=1}^{N}\log_2(1+\gamma_{k,n})}{\sum_{k=1}^{K}P_{c,k}+\sum_{k=1}^{K}\sum_{n=1}^{N}\mu_{k,n}p_{k,n}}\ , \quad (2)$$

wherein $W$ is the subcarrier bandwidth, $\mu_{k,n}$ is the inverse of the efficiency of the $k$-th user's power amplifier over subcarrier $n$ (if a single amplifier is used for all the subcarriers, $\mu_{k,n} = \mu_k$), and $\sum_{k=1}^{K} P_{c,k} = P_c$ is the total static hardware power dissipated to operate all of the network communication links, where $P_{c,k}$ is the static hardware power dissipated by the transceiver of user $k$. It should be noted that (2) is measured in bit-per-Joule, thus naturally representing the amount of information that can be reliably transmitted per Joule of consumed energy.

In this context, the aim of this work is to analyze the problem of power control and subcarrier allocation for GEE maximization, subject to both maximum power and minimum rate constraints. Considering full frequency reuse (i.e. all subcarriers can be assigned to all users), and stacking all the users' powers into the vector $\boldsymbol{p} = \{p_{k,n}\}_{k,n}$, the problem is mathematically stated as

$$\max_{\boldsymbol{p}} \frac{\sum_{k=1}^{K}\sum_{n=1}^{N}\log_2(1+\gamma_{k,n})}{P_c+\sum_{k=1}^{K}\sum_{n=1}^{N}\mu_{k,n}p_{k,n}} \quad (3a)$$

$$\text{s.t.} \sum_{n=1}^{N} p_{k,n} \leq P_{max,k}\ ,\ \forall\, k=1,\ldots,K \quad (3b)$$

$$p_{k,n} \geq 0\ ,\ \forall\, k=1,\ldots,K\ , n=1,\ldots,N \quad (3c)$$

$$\sum_{n=1}^{N}\log_2(1+\gamma_{k,n}) \geq R_{min,k}\ ,\ \forall\, k=1,\ldots,K \quad (3d)$$

where $R_{min,k}$ is a minimum data-rate requirement which has to be guaranteed to each served user. Such a parameter is in general application and/or network-specific, but can not exceed the upper-bound $R_{max,k} = \sum_{n=1}^{N}\lim_{p_{k,n}\to+\infty}\log_2(1+\gamma_{k,n}) = \sum_{n=1}^{N}\log_2(1+\alpha_{k,n}/\xi_{k,n})$. Thus, in the following we assume that $R_{min,k} \leq R_{max,k}$ for all $k \in \mathcal{K}$, since otherwise Problem (3) would be unfeasible. It is also to be observed that Problem (3) performs joint power and subcarrier allocation. Indeed, given that full frequency reuse is considered, the use of a subcarrier is determined by the fact that a non-zero power is allocated over it. Problem (3) is a non-convex fractional problem, and even testing its feasibility would require solving a non-convex feasibility test. However, a sufficient feasibility condition can be obtained by first relaxing (3d) into the per-subcarrier constraint $\log_2(1+\gamma_{k,n}) \geq R_{min,k}^{(n)}$ for all $k$ and $n$, with $\sum_{n=1}^{N} R_{min,k}^{(n)} = R_{min,k}$ for all $k$. Clearly, any power allocation fulfilling the per-subcarrier constraint above also fulfills (3d). Also, the per-subcarrier constraints effectively allow reformulating the feasible set of the relaxed version of (3) into a set of linear inequalities, which enables to derive sufficient feasibility conditions for (3) applying the derivations from [14] to all subcarriers $n = 1,\ldots,N$. The non-convexity of Problem (3) prevents the use of standard convex optimization methods and calls for tailored optimization tools. Fractional programming is the branch of optimization theory which specifically handles fractional problems, and the most widely used fractional programming approach is the Dinkelbach's algorithm [4], [35]. For the case at hand, Dinkelbach's algorithm is stated as shown in Algorithm 1, where $\mathcal{P}$ denotes the feasible set of Problem (3).

**Algorithm 1** Dinkelbach's algorithm for GEE

```
Set j = 0; λ_j = 0; ε > 0;
while F(λ_j) ≥ ε do
```

$$\boldsymbol{p}^\star = \arg\max_{\boldsymbol{p}\in\mathcal{P}} \left\{ \sum_{k=1}^{K}\sum_{n=1}^{N}\log_2(1+\gamma_{k,n}) - \lambda_j\left(P_c+\sum_{k=1}^{K}\sum_{n=1}^{N}\mu_{k,n}p_{k,n}\right)\right\}; \quad (4)$$

$$F(\lambda_j) = \sum_{k=1}^{K}\sum_{n=1}^{N}\log_2(1+\gamma_{k,n}(\boldsymbol{p}^\star)) - \lambda_j\left(P_c+\sum_{k=1}^{K}\sum_{n=1}^{N}\mu_{k,n}p_{k,n}^\star\right); \quad (5)$$

$$\lambda_{j+1} = \frac{\sum_{k=1}^{K}\sum_{n=1}^{N}\log_2(1+\gamma_{k,n}(\boldsymbol{p}^\star))}{P_c+\sum_{k=1}^{K}\sum_{n=1}^{N}\mu_{k,n}p_{k,n}^\star}; \quad j = j+1;$$

**end while**

It is worth noting that Dinkelbach's algorithm tackles the original fractional problem by solving a sequence of non-fractional problems of the form of Problem (4), whose objective is the numerator of (3a) minus the denominator of (3a), weighted by a parameter $\lambda$. Like any instance of Dinkelbach's algorithm, Algorithm 1 enjoys the following two main properties:

- **P1:** Assuming Problem (3) is feasible, then Algorithm 1 yields the global solution of (3) with super-linear convergence rate [35], [36].
- **P2:** The sequence $\{F(\lambda_j)\}_j$ is decreasing and global optimality is attained when $F = 0$.

**Remark 1.** *Properties* **P1** *and* **P2** *hold provided that Problem* (4) *is* globally *solved in each iteration of Algorithm 1. This requirement could be easily fulfilled if* (3a) *had a concave numerator and a convex denominator, as in this case* (4) *would be a convex problem. Unfortunately, due to the presence of multi-user interference,* (3a) *does not possess the concave/convex structure. Furthermore, if* (3) *is not feasible, then an optimal solution does not exist and Algorithm 1 will stop in the first iteration, declaring the unfeasibility of* (3).

Thus, the exact implementation of Algorithm 1 would require an exponential complexity [16]. Even simply testing the feasibility of (3) would lead to a non-convex feasibility test, due to the non-concavity of the QoS constraints in (3d). A more practical approach for GEE maximization has been proposed in [16], by merging sequential optimization and fractional programming. The resulting algorithm, although not being theoretically guaranteed to be globally optimal, has been numerically shown to achieve global optimality in several relevant problem instances. Nevertheless, it still requires to numerically solve a sequence of convex problems, each one having $KN$ optimization variables. In large networks with many users and/or subcarriers, the resulting complexity might still be not practical. Instead, this work will present a new method to tackle (3), which enjoys lower complexity but equal performance as the method in [16]. The approach will be based on a joint use of fractional programming, game theory, and learning theory. First, Section III will address the case in which the rate constraints (3d) are relaxed, and then Section IV will extend the proposed approach to the general case in which rate constraints are enforced. It is worth mentioning that the convergence criterion used in Algorithm 1 is $F(\lambda_j) \geq \epsilon$, with $\epsilon > 0$. The optimal solution is obtained when $\epsilon = 0$, but in general this can be achieved only asymptotically. However, $\epsilon > 0$ allows approaching the optimal solution within any desired accuracy level. Also, the convergence rule in Algorithm 1 is in terms of the auxiliary function $F(\lambda)$, rather than in terms of the variable $\boldsymbol{x}$, as it is done in other contexts [37]. This is typical in iterative algorithms monotonically increasing (decreasing) an objective (cost) function.

**Remark 2.** *The model developed in this section assumes that $\{\alpha_{k,n}\}_{k,n}$, $\{\xi_{k,n}\}_{k,n}$, and $\{\beta_{\ell,n}^{(k)}\}_{k,\ell,n}$ are known to the resource allocator. This holds in a block-fading channel scenario, in which perfect channel estimation is performed at the receiver side. Nevertheless, in real-world scenarios it is possible that the interference coefficients $\{\xi_{k,n}\}_{k,n}$, and $\{\beta_{\ell,n}^{(k)}\}_{k,\ell,n}$ are only partially known. In this case, the algorithms to be developed in the sequel can still be applied in conjunction with robust optimization methods. A viable approach is to reformulate* (3a) *and* (3d) *by considering their expected value with respect to the variables $\{\phi_{k,n}, \beta_{\ell,n}^{(k)}\}$. Since the argument of the expectation is convex in $\{\phi_{k,n}, \beta_{\ell,n}^{(k)}\}$, we can obtain a lower-bound of* (3a) *and* (3d) *by moving the expectations inside their argument, thus maximizing its lower-bound. This yields a formally equivalent problem as* (3), *with $\{\xi_{k,n}\}_{k,n}$ and $\{\beta_{\ell,n}^{(k)}\}_{k,\ell,n}$ are replaced by their mean value. Similarly, if the channels are known up to a maximum estimation error, a robust, max-min approach is to solve* (3) *after replacing $\{\xi_{k,n}\}_{k,n}$ and $\{\beta_{\ell,n}^{(k)}\}_{k,\ell,n}$ by their worst-case values.*

## III. Global energy efficiency maximization without minimum rate constraints

In this section, we tackle (3) when (3d) are removed from the optimization problem. Our departing point is Dinkelback's procedure in Algorithm 1, and we propose to tackle the inner problem (4) modeling it as a potential game, and developing an iterative method which improves the GEE at each iteration and eventually converges towards an efficient solution of (4).

Define the objective of Problem (4) as the function:

$$V(\boldsymbol{p}) = \sum_{i=1}^{K}\sum_{n=1}^{N}\log_2(1+\gamma_{i,n}) - \lambda_j\left(P_c + \sum_{i=1}^{K}\sum_{n=1}^{N}\mu_{i,n}p_{i,n}\right) \tag{6}$$

Since finding the global maximum of (6) is in general computationally prohibitive, the approach will be to provide a computationally efficient method to derive (possibly suboptimal) power allocation vectors $\boldsymbol{p}$. To this end, let us introduce the non-cooperative game in normal form $\mathcal{G} = \{\mathcal{K}, \{\mathcal{S}_k\}_{k=1}^{K}, \{u_k\}_{k=1}^{K}\}$, wherein $\mathcal{K}$ is the players' set,

$$\mathcal{S}_k = \Big\{\boldsymbol{p}_k = [p_{k,1},\ldots,p_{k,N}]^T : p_{k,n} \geq 0,\ \forall n = 1,\ldots,N, \\ \sum_{n=1}^{N} p_{k,n} \leq P_{max,k}\Big\}\,, \tag{7}$$

is the $k$-th player's strategy set, and $u_k$ is the $k$-th player's utility function defined as

$$u_k(\boldsymbol{p}_k, \boldsymbol{p}_{-k}) = V(\boldsymbol{p}_k, \boldsymbol{p}_{-k}) = V(\boldsymbol{p}) \tag{8}$$

where $\boldsymbol{p}_{-k} = \{\boldsymbol{p}_j\}_{j\in\mathcal{K}, j\neq k}$. In particular, it can be seen that every player has the same utility function $V(\boldsymbol{p})$, which makes $\mathcal{G}$ a potential game [38], and specifically a so-called *identical interest* game [39]. Potential games enjoy several pleasant properties, which makes them attractive for wireless resource allocation [40], [41]. In particular, potential games guarantee the existence of at least one pure-strategy NE, under the mild assumptions that the potential function is continuos and the strategy sets are compact [38]. Moreover, pure-strategy NE can be reached by implementing the game best-response dynamics, i.e. letting each player $k$ iteratively maximize the common utility function $V(\boldsymbol{p})$ with respect to their own strategy $\boldsymbol{p}_k$, assuming all other players' power vectors $\{\boldsymbol{p}_\ell\}_{\ell\neq k}$ are fixed. Unfortunately, implementing the best-response dynamics of $\mathcal{G}$ is no easy task, since the common utility $V(\boldsymbol{p})$ is not concave even with respect to only a single power vector $\boldsymbol{p}_k$. To circumvent this problem, in the following we replace the concept of best-response, with the milder notion of better response [42]:

**Definition 1** (Better-Response Strategy)**.** *$\boldsymbol{p}_k^*$ is a better-response strategy for user $k \in \mathcal{K}$ to $(\boldsymbol{p}_k, \boldsymbol{p}_{-k})$, if $u_k(\boldsymbol{p}_k^*, \boldsymbol{p}_{-k}) \geq u_k(\boldsymbol{p}_k, \boldsymbol{p}_{-k})$, i.e. $\boldsymbol{p}_k^*$ dominates $\boldsymbol{p}_k$ when other players choose $\boldsymbol{p}_{-k}$.*

Otherwise stated, each player $k$ does not aim at computing the strategy which maximizes its utility function $u_k$ given the strategies of the other players. Instead, the goal is just to find a strategy $\boldsymbol{p}_k^*$ which increases $u_k$ compared to the present strategy $\boldsymbol{p}_k$, and given the strategies of the other players $\boldsymbol{p}_{-k}$. This introduces the notion of better-response dynamics [42]:

**Definition 2** (Better-Response Dynamics (BRD))**.** *Let $\boldsymbol{p}(i)$ denote the $KN \times 1$ vector $[\boldsymbol{p}_1^T(i),\ldots,\boldsymbol{p}_K^T(i)]^T$ collecting the strategies played by the $K$ players at time $i$. Then, a Better-Response Dynamics is a sequence of strategies $\{\boldsymbol{p}^*(i)\}_i$, wherein, at each time instant $i$, and for all $k$,* $\boldsymbol{p}_k^*(i)$ *is a better response of player k, to the strategies*[1] $[\boldsymbol{p}_1^T(i), \ldots, \boldsymbol{p}_{k-1}^T(i), \boldsymbol{p}_{k+1}^T(i-1), \ldots, \boldsymbol{p}_K^T(i)]^T$.

**Proposition 1.** *Let* $\{\boldsymbol{p}^*(i)\}_i$ *be a better-response dynamics for the game* $\mathcal{G}$. *Then, the sequence* $\{V(\boldsymbol{p}^*(i))\}_i$ *is monotonically convergent, i.e.* $\lim_{i\to\infty}(V(\boldsymbol{p}^*(i)) - V(\boldsymbol{p}^*(i-1))) = 0$.

*Proof:* According to the definition of better-response strategy, any unilateral strategy update performed by user $k \in \mathcal{K}$ improves the utility function, and thus the potential function. So, $V(\boldsymbol{p}^*(i)) \geq V(\boldsymbol{p}^*(i-1)$ for any $i = 1, 2, \ldots, \infty$, which shows how $V(\boldsymbol{p}^*(i))$ is monotonically increasing. Moreover, $V(\boldsymbol{p})$ admits a finite maximum over the set $\mathcal{S} = \mathcal{S}_1 \cap \mathcal{S}_2 \ldots \cap \mathcal{S}_K$, due to Weierstrass extreme value theorem. Indeed, $V(\boldsymbol{p})$ is continuous and $\mathcal{S}_k$ is compact for each $k$. Hence, $V(\boldsymbol{p})$ can not grow indefinitively over the feasible set, and this shows the thesis. ■

We stress that Proposition 1 holds without requiring any concavity/convexity property for the potential $V(\boldsymbol{p})$ and for the strategy sets $\mathcal{S}_k$. Moreover, it is important to remark that while the condition $\lim_{i\to\infty}(V(\boldsymbol{p}^*(i)) - V(\boldsymbol{p}^*(i-1))) = 0$ can only be achieved asymptotically, the monotonic increasing behavior of $V(\boldsymbol{p}(i))$ enables to approach it within any desired tolerance $\epsilon$. Practically speaking, convergence is declared when no further significant improvement can be obtained after two consecutive iterations, i.e. when $V(\boldsymbol{p}^*(i)) - V(\boldsymbol{p}^*(i-1) \leq \epsilon$, which is achieved in a finite amount of iterations for any $\epsilon > 0$. Finally, we would like to remark that, though convergence of the BRD to an efficient solution is guaranteed by Proposition 1, convergence towards a NE is possible but cannot be guaranteed in our case. This is because the NE concept relies on the definition of best response functions which, in our case, cannot be computed due to the non-convexity of the best-response problems.

Next, being guaranteed that the better-response dynamics will converge, it remains to devise an efficient method to compute the players' better responses. To this end, let us observe that, after some elaborations, the utility function of player $k$ can be rewritten as follows:

$$u_k(\boldsymbol{p}_k, \boldsymbol{p}_{-k}) = \sum_{n=1}^{N} u_{k,n}(p_{k,n}, \boldsymbol{p}_{-k}) \tag{9}$$

where

$$\begin{aligned} &u_{k,n}(p_{k,n}, \boldsymbol{p}_{-k}) = \log_2(1+\gamma_{k,n}) \\ &+\sum_{i\neq k}\log_2\left(\sigma_n^2+(\alpha_{i,n}+\xi_{i,n})p_{i,n}+\beta_{k,n}^{(i)}p_{k,n}+\sum_{\ell\neq i, l\neq k}\beta_{\ell,n}^{(i)}p_{\ell,n}\right) \\ &-\sum_{i\neq k}\log_2\left(\sigma_n^2+\xi_{i,n}p_{i,n}+\beta_{k,n}^{(i)}p_{k,n}+\sum_{\ell\neq i, l\neq k}\beta_{\ell,n}^{(i)}p_{\ell,n}\right) \\ &-\lambda_j\left(\frac{P_c}{N}+\mu_{k,n}p_{k,n}\right)-\lambda_j\sum_{\ell\neq k}\mu_{\ell,n}p_{\ell,n} \end{aligned} \tag{10}$$

[1]Without loss of generality, we assume that the players play one after the other and are indexed in the order in which they play. Nevertheless, the analysis and convergence results to follow can be straightforwardly extended to the scenario in which the users perform asynchronous power updates.

Let us now denote by $\bar{\boldsymbol{p}}_k$ the current strategy of player $k$, and observe that the non-concavity of (10) emerges from the third term in (10). Thus, a convenient way of finding a better response for player $k$ is to consider a concave lower-bound of (10) for all $n = 1, \ldots, N$, obtained by replacing the non-concave part by its first-order Taylor expansion around $\bar{\boldsymbol{p}}_k$, namely[2]:

$$\hat{u}_k(p_{k,n}, \boldsymbol{p}_{-k}, \bar{\boldsymbol{p}}_k) = \textstyle\sum_{n=1}^{N} \hat{u}_{k,n}(p_{k,n}, \boldsymbol{p}_{-k}, \bar{p}_{k,n}) \tag{11}$$

with

$$\begin{aligned} &\hat{u}_{k,n}(p_{k,n}, \boldsymbol{p}_{-k}, \bar{p}_{k,n}) = \log_2(1+\gamma_{k,n}) + \phi_{k,n}(p_{k,n}-\bar{p}_{k,n}) \\ &+\sum_{i\neq k}\log_2\left(\sigma_n^2+\eta_{i,n}p_{i,n}+\beta_{k,n}^{(i)}p_{k,n}+\sum_{\ell\neq i,\ell\neq k}\beta_{\ell,n}^{(i)}p_{\ell,n}\right) \\ &-\textstyle\sum_{i\neq k}\log_2\left(\sigma_n^2+\xi_{i,n}p_{i,n}+\beta_{k,n}^{(i)}\bar{p}_{k,n}+\sum_{\ell\neq i,\ell\neq k}\beta_{\ell,n}^{(i)}p_{\ell,n}\right) \\ &-\lambda_j\left(\frac{P_c}{N}+\mu_{k,n}p_{k,n}-\textstyle\sum_{\ell\neq k}\mu_{\ell,n}p_{\ell,n}\right) \end{aligned} \tag{12}$$

and

$$\eta_{k,n} = \alpha_{k,n} + \xi_{k,n} \tag{13}$$

$$\phi_{k,n} = -\sum_{i\neq k}\frac{\beta_{k,n}^{(i)}/\log(2)}{\sigma_n^2+\xi_{i,n}p_{i,n}+\beta_{k,n}^{(i)}\bar{p}_{k,n}+\sum_{\ell\neq i,\ell\neq k}\beta_{\ell,n}^{(i)}p_{\ell,n}} \tag{14}$$

The approximate function $\hat{u}_k$ allows determining a better response for player $k$, as proved by the following Proposition 2.

**Proposition 2.** *Consider the concave optimization problem:*

$$\max_{\boldsymbol{p}_k\in\mathcal{S}_k} \hat{u}_k(\boldsymbol{p}_k, \boldsymbol{p}_{-k}, \bar{\boldsymbol{p}}_k) \tag{15}$$

*Problem* (15) *admits a unique solution, which is also a better-response strategy for user* $k \in \mathcal{K}$.

*Proof:* See Appendix VIII-A. ■

So far, we have shown that the better response dynamics of $\mathcal{G}$ converges, and we have derived a convex problem that enables computing the better response for each player $k$. Nevertheless, employing standard convex optimization methods to solve (15) would require polynomial complexity in the number of subcarriers $N$, since (15) has $N$ optimization variables. This might still be impractical for high $N$. Instead, a faster approach with guaranteed linear complexity is obtained resorting to learning theory, as shown in the following proposition.

**Proposition 3.** *Consider the following iterative learning mechanism with exponential mappings:*

$$\begin{cases} y_{k,n}(t+1) = y_{k,n}(t) + \delta_t \hat{v}_{k,n}(p_{k,n}(t), \boldsymbol{p}_{-k}, \bar{\boldsymbol{p}}_k) \\ p_{k,n}(t+1) = P_{max,k}\frac{e^{y_{k,n}(t)}}{1+\sum_{m=1}^{N}e^{y_{k,m}(t)}} \end{cases} \tag{16}$$

*where* $t$ *is the iteration index of the learning procedure,* $\delta_t$ *is the step-size,* $\bar{\boldsymbol{p}}_k$ *is the strategy of user* $k$ *at the previous*

[2]From the necessary and sufficient first-order convexity condition [43], for any $\boldsymbol{x}$, $\boldsymbol{y}$ in the domain of $f$, it holds: $f(\boldsymbol{x}) \geq f(\boldsymbol{y}) + (\nabla f(\boldsymbol{y}))^T(\boldsymbol{x}-\boldsymbol{y})$, and thus $f(\boldsymbol{x})$ is lower-bounded by its first-order Taylor expansion around $\boldsymbol{y}$. The term that is linearized in (10) is convex, and therefore it is lower-bounded by its Taylor expansion around any point.

*iteration of the BRD. If* $\sum_{t=1}^{+\infty} \delta_t^2 < \sum_{t=1}^{+\infty} \delta_t = +\infty$*, Iteration* (16) *is such that* $\lim_{t\to\infty} \|\boldsymbol{p}_k(t) - \boldsymbol{p}_k(t-1)\| = 0$*, with* $\boldsymbol{p}_k(t) = \{p_{k,n}(t)\}_{n=1}^N$*, and the limit point* $\tilde{\boldsymbol{p}}_k$ *obtained upon convergence is the unique solution of Problem* (15).

*Proof:* See Appendix VIII-B. ■

The proposed learning mechanism in (16) enjoys several interesting properties. First, it is *reinforcing* as the so-called *scores* $y_{k,n}$ are updated according to the *marginal utility* $\hat{v}_{k,n}$, defined as the first order derivative of (12) with respect to $p_{k,n}(i)$, namely:

$$\hat{v}_{k,n}(p_{k,n}(t), \boldsymbol{p}_{-k}, \bar{\boldsymbol{p}}_k) = \frac{1}{\log(2)}\Bigg[ \tag{17}$$

$$\frac{\eta_{k,n}}{\sigma_n^2+\eta_{k,n}p_{k,n}(t)+\sum_{i\neq k}\beta_{i,n}^{(k)}p_{i,n}} - \frac{\xi_{k,n}}{\sigma_n^2+\xi_{k,n}p_{k,n}(t)+\sum_{i\neq k}\beta_{i,n}^{(k)}p_{i,n}}$$

$$+\sum_{i\neq k}\frac{\beta_{k,n}^{(i)}}{\sigma_n^2+\eta_{i,n}p_{i,n}+\beta_{k,n}^{(i)}p_{k,n}(t)+\sum_{l\neq i,l\neq k}\beta_{l,n}^{(i)}p_{l,n}}\Bigg] - \nu_{k,n}, \tag{18}$$

where $\nu_{k,n} = \lambda_j \mu_{k,n} - \phi_{k,n}$. Also, the exponential mapping relates each score $y_{k,n}$ to a feasible power level $p_{k,n}$, and always generate solutions simultaneously satisfying all problem constraints. In fact, the power levels generated by (16) satisfy $\sum_{n\in\mathcal{N}} p_{k,n}(t) \leq P_{max,k}$ for all $t$ and $k$.

Based on the derived results, a better response for player $k$ can be computed as shown in Algorithm 2, while the overall GEE maximization procedure can be stated as in Algorithm 3.

**Algorithm 2** Unilateral better-response for User $k$

```
Set t = 0; y_{k,n} = 0;
while Convergence is not achieved do
    t = t + 1;
    for each n = 1, ..., N do simultaneously
        p_{k,n} = P_{max,k} e^{y_{k,n}} / (1 + sum_{m=1}^{N} e^{y_{k,m}});
        y_{k,n} = y_{k,n} + delta_t v̂_{k,n};
    end for
end while
```

**Algorithm 3** GEE maximization without QoS constraints

Set $j = 0$; $\lambda_j = 0$; $\varepsilon > 0$;
**while** $\bar{F}(\lambda_j) \geq \varepsilon$ || $|\lambda_j - \lambda_{j-1}| \geq \varepsilon$ **do**
  **while** Convergence has not been reached **do**
    **for each** $k = 1, \ldots, K$ **do**
      $\boldsymbol{p}_k^* \leftarrow$ The solution of Algorithm 2;
      $\bar{\boldsymbol{p}} = (\boldsymbol{p}_k^*, \boldsymbol{p}_{-k})$;
    **end for**
  **end while**

$$\bar{F}(\lambda_j)=\sum_{k=1}^{K}\sum_{n=1}^{N}\log_2(1+\gamma_{k,n}(\bar{\boldsymbol{p}}))-\lambda_j\left(\sum_{k=1}^{K}P_{c,k}+\sum_{n=1}^{N}\mu_{k,n}\bar{p}_{k,n}\right) \tag{19}$$

$$\lambda_{j+1} = \frac{\sum_{k=1}^K \sum_{n=1}^N \log_2(1+\gamma_{k,n}(\bar{\boldsymbol{p}}))}{\sum_{k=1}^K P_{c,k} + \sum_{n=1}^N \mu_{k,n}\bar{p}_{k,n}}; \quad j = j+1$$

**end while**

Let us note that Algorithm 3 is a low-complexity implementation of Algorithm 1 in which the convergence point of the better-response dynamics is used in place of the optimal solution of the auxiliary NP-hard Problem (4). In Algorithm 3, the notation $\overline{F}(\lambda)$ is used to stress the difference with respect to $F(\lambda)$. Specifically, $\overline{F}(\lambda)$ denotes the value of the objective of (4) evaluated for $\boldsymbol{p} = \bar{\boldsymbol{p}}$, with $\bar{\boldsymbol{p}}$ being the power allocation obtained upon convergence of the better response dynamics. Instead, $F(\lambda)$ is the maximum value of the objective of (4). In general it must hold $\overline{F} \leq F$. Thus, recalling the discussion in Section II, the convergence and optimality of Algorithm 3 do not follow from the known properties of Dinkelbach's algorithm. Nevertheless, while Algorithm 3 is not guaranteed to be optimal, its convergence can be theoretically guaranteed. To this end, it is important to observe that the convergence rule of Algorithm 3 is slightly different from that of Algorithm 1, checking the convergence both in terms of $\{\bar{F}(\lambda_j)\}$ and $\{\lambda_j\}_j$. Keeping this in mind, the following result holds.

**Proposition 4.** *Algorithm 3 converges in a finite number of iterations. Moreover, one of the following two cases occurs:*

1) *Algorithm 3 monotonically increases the value of* (3a) *after each iteration and converges. In addition, if upon convergence it holds* $F(\lambda) = 0$*, then global optimality is attained.*
2) *Let* $\bar{j}$ *be the index of the first iteration for which* $\bar{F}(\lambda_j) < 0$*. In this case, Algorithm 3 stops at iteration* $\bar{j}$*, after having monotonically increased the value of* (3a) *for all* $0 \leq j \leq \bar{j}$*.*

*Proof:* See Appendix VIII-C. ■

To execute Algorithm 3, the coefficients $\{\alpha_{k,n}\}_{k,n}$, $\{\xi_{k,n}\}_{k,n}$, and $\{\beta_{\ell,n}^{(k)}\}_{k,\ell,n}$ must be available, which requires information sharing among the $J$ receivers, e.g., the base stations (BSs). This can be achieved through different approaches, but here we identify Cooperative Multi-Point (CoMP) [44]–[46] as a suitable and effective paradigm to implement Algorithm 3 in a centralized fashion. According to the CoMP paradigm, all base stations share CSI with one base station, which acts as head of the cluster. Then, Algorithm 3 is locally executed at the cluster head, and then the resulting resource allocation is shared with the other base stations, which in turn inform their respective mobile users.

## IV. GEE maximization with minimum rate constraints.

This section tackles the general case of Problem (3), in which also the minimum rate constraints in (3d) are enforced. The main difficulty of this scenario lies in the fact that the presence at the same time of maximum power constraints and minimum rate requirements might make the problem unfeasible. In the rest of this section two approaches will be developed. The former will employ a barrier-based reformulation of (3d), whereas the latter will resort to the theory of generalized game theory. In both cases, our departure point is again Algorithm 1, where now the set $\mathcal{P}$ denotes the feasible set of (3), also including the rate constraints in (3d).

### A. Barrier method

The idea of the barrier method is to reformulate the auxiliary problem (4) in Algorithm 1, embedding the rate constraints

$$\phi^{I}_{k,n}=\frac{\rho}{\log_2^2}\left[\frac{\frac{\eta_{k,n}}{\sigma_n^2+\eta_{k,n}p^*_{k,n}+\sum_{i\neq k}\beta^{(k)}_{i,n}p_{i,n}}-\frac{\xi_{k,n}}{\sigma_n^2+\xi_{k,n}p^*_{k,n}+\sum_{i\neq k}\beta^{(k)}_{i,n}p_{i,n}}}{\sum_{m=1}^{N}\log_2\left(1+\frac{p^*_{k,n}\alpha_{k,n}}{\sigma_n^2+\xi_{k,n}p^*_{k,n}+\sum_{\ell\neq k}p_{\ell,n}\beta^{(k)}_{\ell,n}}\right)-R_{min,k}}\right] \tag{20}$$

$$\phi^{II}_{k,n}=\frac{\rho}{\log_2^2}\sum_{i\neq k}\left[\frac{\frac{\beta^{i}_{k,n}}{\sigma_n^2+\eta_{i,n}p_{i,n}+\sum_{l\neq k,l\neq i}p_{l,n}\beta^{(i)}_{l,n}+\beta^{(i)}_{k,n}p^*_{k,n}}-\frac{\beta^{i}_{k,n}}{\sigma_n^2+\xi_{i,n}p_{i,n}+\sum_{l\neq k,l\neq i}p_{l,n}\beta^{(i)}_{l,n}+\beta^{(i)}_{k,n}p^*_{k,n}}}{\sum_{m=1}^{N}\log_2\left(1+\frac{p_{i,m}\alpha_{i,m}}{\sigma_m^2+\xi_{i,m}p_{i,m}+\sum_{l\neq k,l\neq i}p_{l,m}\beta^{(i)}_{l,m}+\beta^{(i)}_{k,m}p^*_{k,m}}\right)-R_{min,i}}\right] \tag{21}$$

into the objective function. To this end, let us define the logarithmic barrier function

$$\varrho(\boldsymbol{p})=\rho\sum_{k=1}^{K}\varrho_k(\boldsymbol{p})=\rho\sum_{k=1}^{K}\log_2\left(\sum_{n=1}^{N}\log_2(1+\gamma_{k,n})-R_{min,k}\right) \tag{22}$$

with $\rho$ a positive cost parameter, and then reformulate (4) as

$$\max_{\mathcal{S}} V(\boldsymbol{p})+\varrho(\boldsymbol{p}) \;, \tag{23}$$

wherein $V(\boldsymbol{p})$ is given by (6), and $\mathcal{S}=\mathcal{S}_1\times\ldots\times\mathcal{S}_K$, with $\mathcal{S}_k$ given by (7) for all $k=1,\ldots,K$. Thus, (23) has been obtained from the original auxiliary problem (4) by relaxing the QoS constraints, but adding the penalty term $\varrho(\boldsymbol{p})$ to the objective function, which ensures that the QoS constraints are satisfied. Indeed, $\varrho(\boldsymbol{p})\to-\infty$ whenever a user's rate tends towards its minimum acceptable rate. Moreover, the pricing parameter $\rho$ weighs the relative importance between the original objective $V(\boldsymbol{p})$ and the barrier term $\varrho(\boldsymbol{p})$. It is also interesting to observe that such a parametric logarithmic barrier method is the typical approach used by the popular interior-point method to solve constrained convex optimization problems [43].

Next, we proceed as in Section III to tackle (23), introducing the potential function

$$V^R(\boldsymbol{p})=V(\boldsymbol{p})+\varrho(\boldsymbol{p}) \;, \tag{24}$$

and defining an identical interest game $\mathcal{G}^R$, where the shared utility function is given by the potential $V^R(\boldsymbol{p})$ in (24). A better response for any $k\in\mathcal{K}$ can be found by following the same steps as in Section III. However, a major difference compared to Section III is that the original problem might not be feasible, i..e, it is possible that no $\boldsymbol{p}_k$ exists that guarantees a finite value of the barrier function $\varrho(\boldsymbol{p})$. This can be determined by solving the following feasibility test:

$$\max_{\boldsymbol{p}_k} \; 1 \tag{25}$$

$$\text{s.t.} \; \sum_{n=1}^{N}p_{k,n}\leq P_{max,k} \tag{26}$$

$$\sum_{n=1}^{N}\log_2\left(1+\frac{p_{k,n}\alpha_{k,n}}{\sigma_n^2+\xi_{k,n}p_{k,n}+\sum_{\ell\neq k}p_{\ell,n}\beta^{(k)}_{\ell,n}}\right)\geq R_{min,k} \;. \tag{27}$$

The test in (25) is convex since the rate function can be seen to be concave with respect to the powers $\{p_{k,n}\}_{n=1}^{N}$, and thus can be efficiently solved. If the test result is negative, no better response for user $k$ exists since its rate constraint can not be fulfilled[3], and user $k$ must either accept a lower rate, or refrain from transmitting in the present channel coherence block. If instead, the test result is positive, then, following similar steps as in Section III, a better response of player $k$ can be determined as the solution of the problem

$$\max_{\boldsymbol{p}_k\in\mathcal{S}_k} \; \hat{u}^R_k(\boldsymbol{p}_k,\boldsymbol{p}_{-k},\boldsymbol{p}^*_k) \tag{28}$$

wherein

$$\hat{u}^R_k(\boldsymbol{p}_k,\boldsymbol{p}_{-k},\boldsymbol{p}^*_k)=\hat{u}_k(\boldsymbol{p}_k,\boldsymbol{p}_{-k},\boldsymbol{p}^*_k)+\sum_{n=1}^{N}(\phi^{I}_{k,n}+\phi^{II}_{k,n})p_{k,n} \tag{29}$$

with $\hat{u}_k(\boldsymbol{p}_k,\boldsymbol{p}_{-k},\boldsymbol{p}^*_k)$,$\phi^{I}_{k,n}$ and $\phi^{II}_{k,n}$ being defined in (11), (20) and (21), respectively.

Problem (28) can be solved by means of the learning mechanism (16). Thus, in the case in which all better responses are feasible, an implementation of Algorithm 1 can be obtained by considering the solution obtained upon convergence of the better-response dynamics in place of the exact solution of the auxiliary problem in each iteration of Algorithm 1. The resulting algorithm is formulated below as Algorithm 4, and enjoys similar properties as Algorithm 3.

**Algorithm 4** GEE maximization with QoS: Barrier method

Set $j=0$; $\lambda_j=0$; $\varepsilon>0$;
**while** $\bar{F}(\lambda_j)\geq\varepsilon \;||\; |\lambda_j-\lambda_{j-1}|\geq\varepsilon$ **do**
  **while** Better response dynamics has not converged **do**
    **for each** $k=1,\ldots,K$ **do**
      $\boldsymbol{p}^*_k\leftarrow$ The solution of Problem (28) through Algorithm 2;
      $\bar{\boldsymbol{p}}=(\boldsymbol{p}^*_k,\boldsymbol{p}_{-k})$;
    **end for**
  **end while**

$$\bar{F}(\lambda_j)=\sum_{k=1}^{K}\sum_{n=1}^{N}\log_2(1+\gamma_{k,n}(\bar{\boldsymbol{p}}))-\lambda_j\left(\sum_{k=1}^{K}P_{c,k}+\sum_{n=1}^{N}\mu_{k,n}\bar{p}_{k,n}\right)$$

$$\lambda_{j+1}=\frac{\sum_{k=1}^{K}\sum_{n=1}^{N}\log_2(1+\gamma_{k,n}(\bar{\boldsymbol{p}}))}{\sum_{k=1}^{K}P_{c,k}+\sum_{n=1}^{N}\mu_{k,n}\bar{p}_{k,n}}; \quad j=j+1;$$

**end while**

[3]We stress that such circumstance is not related to the radio resource allocation policy, but rather to the deployment of the system, to the contingent channel conditions, and to the parameters $P_{max,k}$ and $R_{min,k}$.

### B. Generalized games method

The approach from Section IV-A requires to solve the feasibility test (25) before computing each player's better response. Although convex, solving (25) produces an additional complexity which might not be desirable, especially for large $K$ and $N$. In order to further reduce the complexity burden, here we propose to reformulate the per-user constraints in (3d), into per-subcarrier QoS constraints, namely considering, for all $k = 1, \ldots, K$, the set of $N$ constraints

$$\log_2 (1 + \gamma_{k,n}) \geq R_{min,k}^{(n)} \ , \ \forall \ n = 1, \ldots, N \ , \qquad (30)$$

with $\sum_{n=1}^{N} R_{min,k}^{(n)} = R_{min,k}$. Clearly, every power allocation that fulfills (30), also satisfies (3d) but the reverse statement is not true. As a consequence, the reformulation in (30) might cause a slight performance degradation, but, as shown in the sequel, it allows for a much less complex power allocation algorithm. Before addressing this point and delving into the mathematical details of the allocation procedure, it is worth stressing that a per-subcarrier power constraint is in line with the IEEE 802.11 standard for WiFi systems, which enforces equal target rates over each subcarrier, as also mentioned in the recent work [47]. Moreover, as for the choice of each $R_{min,k}^{(n)}$ in (30) we make the following remark.

**Remark 3.** *In order to avoid unfeasibility due to the presence of only a few subcarriers with poor channel conditions, $R_{min,k}^{(n)}$ should be chosen in a subcarrier-dependent way. In particular, since the rate over each subcarrier is upper-bounded by $R_{max,k}^{(n)} = \log_2 \left(1 + \frac{\alpha_{k,n}}{\xi_{k,n}}\right)$, it is natural to choose $R_{min,k}^{(n)}$ as a fraction of $R_{max,k}^{(n)}$ for each $n = 1, \ldots, N$.*

Next, having reformulated the QoS as in (30) for all $k = 1, \ldots, K$, we again consider the potential function in (6), and define the non-cooperative potential game $\mathcal{G}_{QoS} = \{\mathcal{K}, \{\mathcal{S}_k\}_{k=1}^K, \{u_k\}_{k=1}^K\}$, wherein $\mathcal{K}$ is the players' set, $u_k$ is the $k$-th player's utility function defined as $u_k = V(\boldsymbol{p})$, and $\mathcal{S}_k$ is the $k$-th player's strategy set, which is now defined to account for the presence of minimum rate constraints, namely:

$$\mathcal{S}_k(\boldsymbol{p}_{-k}) = \Big\{ \boldsymbol{p}_k \in \mathcal{S}_k : \log_2 \left(1 + \gamma_{k,n}(\boldsymbol{p}_k, \boldsymbol{p}_{-k})\right) \geq R_{min,k}^{(n)}, \ \forall n = 1, \ldots, N \Big\}. \qquad (31)$$

In (31), the notation $\mathcal{S}_k(\boldsymbol{p}_{-k})$ is used to stress the fundamental fact that when QoS requirements are enforced, the strategy set of player $k$ depends on the strategies of the other players $\boldsymbol{p}_{-k}$. Otherwise stated, in the considered game, not only the utility functions, but also the players' strategy sets are coupled. This is the defining property of the so-called *generalized* non-cooperative games [13]–[15], whose analysis is more involved than for regular non-cooperative games. In particular, the convergence of the best/better-response dynamics is more difficult to prove, mainly due to the fact that the generic best/better-response problem might be unfeasible.

To elaborate, for the case at hand it is still convenient to consider the better response dynamics of $\mathcal{G}_{QoS}$, as in the case without QoS constraints. Following similar steps as in Section III, we obtain that a better response for player $k$ can be computed as the solution of a problem that is formally equivalent to Problem (15), with the addition of QoS constraints, namely

$$\max_{\boldsymbol{p}_k \in \mathcal{S}_k} \ \hat{u}_k(\boldsymbol{p}_k, \boldsymbol{p}_{-k}, \bar{\boldsymbol{p}}_k) \qquad (32a)$$

$$\text{s.t.} \quad \log_2(1 + \gamma_{k,n}) \geq R_{min,k}^{(n)} \ , \forall \ n = 1, \ldots, N \qquad (32b)$$

At this point we exploit the fact that per-subcarrier QoS constraints have been enforced. Indeed, unlike Section IV-A where the feasibility of a user's rate requirement had to be determined by solving the test (25), here a closed-form feasibility condition can be derived to test the feasibility of (32). Also, observe that (32b) can be equivalently reformulated, for all $n = 1, \ldots, N$, as

$$p_{k,n} \geq \frac{(2^{R_{min,k}^{(n)}} - 1)\left(\sigma^2 + \sum_{\ell \neq k} p_{\ell,n} \beta_{\ell,n}^{(k)}\right)}{\alpha_{k,n} - \xi_{k,n}(2^{R_{min,k}^{(n)}} - 1)} = P_{min,k}^{(n)} \ , \qquad (33)$$

and therefore (32) is feasible if and only if the following condition holds:

$$\sum_{n=1}^{N} P_{min,k}^{(n)} \leq P_{max,k} \ . \qquad (34)$$

Let $\Delta_k = P_{max,k} - \sum_{n=1}^{N} P_{min,k}^{(n)}$ be the residual power for each player $k$. Then, assuming the problem is feasible, i.e., $\Delta_k > 0$, and exploiting the fact that (32b) is equivalent to the linear constraint (33), it is possible to solve (32) as shown in the next proposition.

**Proposition 5.** *Consider the following iterative learning mechanism with exponential mappings:*

$$\begin{cases} y_{k,n}(t+1) = y_{k,n}(t) + \delta_t \hat{v}_{k,n}(p_{k,n}(t), \boldsymbol{p}_{-k}, \bar{\boldsymbol{p}}_k) \\ p'_{k,n}(t+1) = \Delta_k \frac{e^{y_{k,n}(t)}}{1 + \sum_{m=1}^{N} e^{y_{k,m}(t)}} \\ p_{k,n}(t+1) = p'_{k,n}(t+1) + P_{min,k}^{(n)} \end{cases} \qquad (35)$$

*where $t$ is the iteration index of the learning procedure, $\delta_t$ is the step-size, $\bar{\mathbf{p}}_k$ is the strategy of user $k$ at the previous iteration of the BRD, and $\hat{v}_{k,n}(p_{k,n}(t), \boldsymbol{p}_{-k}, \bar{\boldsymbol{p}}_k)$ is defined in* (17). *If $\sum_{t=1}^{+\infty} \delta_t^2 < \sum_{t=1}^{+\infty} \delta_t = +\infty$,* (35) *converges to the unique solution of Problem* (32).

*Proof:* See Appendix VIII-D. ■

It is worth remarking that the learning mechanism (35) is similar to that in (16). The only difference between the two mechanisms is that (35) is defined over the shrunken feasible set $\prod_{n=1}^{N} [0, \Delta_k]$, and thus requires an additional equation to guarantee the feasibility of the solutions.

Finally, an implementation of Algorithm 1 can be developed as shown in Algorithm 5.

It is worth noting that if the feasibility condition in (33) does not hold for some users and subcarriers, the fact that per-subcarrier QoS constraints have been considered, enables to relax the rate requirement only for the specific subcarriers which experience poor channel conditions. That is, the minimum transmission power constraint is enforced on a subset $\mathcal{N}_k^F \subseteq \mathcal{N}$ such that $\sum_{n \in \mathcal{N}_k^F} P_{min,k}^{(n)} \leq P_{max,k}$, while the

**Algorithm 5** GEE maximization with QoS: Generalized games method

```
Set j = 0; λ_j = 0; ε > 0; Δ_k = P_{max,k} − Σ_{n=1}^{N} P^{(n)}_{min,k}, ∀k ∈ 𝒦
while F̄(λ_j) ≥ ε || |λ_j − λ_{j−1}| ≥ ε do
    while Better response dynamics has not converged do
        for each k = 1, ..., K do
            p'_k ← The solution of Algorithm 2 with P_{max,k} = Δ_k;
            p*_k ← {p'_{k,n} + P^{(n)}_{min,k}}_n;
            p̄ = (p*_k, p_{−k});
        end for
    end while
```

$$\bar{F}(\lambda_j)=\sum_{k=1}^{K}\sum_{n=1}^{N}\log_2(1+\gamma_{k,n}(\bar{\boldsymbol{p}}))-\lambda_j\left(\sum_{k=1}^{K}P_{c,k}+\sum_{n=1}^{N}\mu_{k,n}\bar{p}_{k,n}\right) \tag{36}$$

$$\lambda_{j+1}=\frac{\sum_{k=1}^{K}\sum_{n=1}^{N}\log_2(1+\gamma_{k,n}(\bar{\boldsymbol{p}}))}{\sum_{k=1}^{K}P_{c,k}+\sum_{n=1}^{N}\mu_{k,n}\bar{p}_{k,n}};\quad j=j+1;$$

**end while**

constraint is removed from the remaining subcarriers, i.e., $P_{min,k}^{(n)}=0$ if $n\notin\mathcal{N}_k^F$. This is in line with the well-known water-filling solution for a set of parallel channels, in which low power is allocated to channels with poor propagation conditions.

## V. Computational complexity analysis

### A. Asymptotic complexity of Algorithm 3

Algorithm 3 is composed of three nested loops. The inner loop is Algorithm 2, which is used to compute the users' better responses. The middle loop is the better response dynamics, while the outer loop is the update of $\lambda$ according to Dinkelbach's rule.

Algorithm 2 requires to implement the learning scheme (16) until convergence. In each iteration of Algorithm 2, the variables $p_{k,n}$ and $y_{k,n}$ are simultaneously updated for each of the $N$ subcarriers. Each update requires computing formula (16), which has complexity $\mathcal{O}(1)$. Thus, the per-iteration complexity of Algorithm 2 is $\mathcal{O}(N)$. On the other hand, no closed-form result is available for the number $I_L$ of iterations before Algorithm 2 converges, even though the numerical analysis presented in Section VI shows that a few tens of iterations are required.

As for the middle loop, the number of better responses to compute until convergence will be equal to the number of users $K$, times the per-user amount of iterations, say $I_B$. Unfortunately, a closed-form expression which quantifies $I_B$ is not available, even though related works on game-theoretic resource allocation algorithms show that typically $I_B$ is of the order of a few units [7], [14], [48]. This behavior has been confirmed by our numerical analysis, too.

As for the number of iterations required for the convergence of the outer loop, say $I_D$, again no closed-form result is available in general, even though, as observed in Section II, the convergence rate of Dinkelbach's algorithm is super-linear. This argument is corroborated by our numerical results, which shows how $I_D$ is of the order of a few units. Also, assuming upper and lower bounds of the maximum GEE value are available, say $U$ and $L$, we could find the zero of $F(\lambda)$ by updating $\lambda$ according to the bisection method, instead of using Dinkelbach's update rule. Although bisection converges typically slower than Dinkelbach's method [47], it provides an estimate of $I_D$. Upon using the bisection method, the zero of $F(\lambda)$ can be found within a tolerance $\epsilon$ with $I_D=\log_2\left(\left\lceil\frac{U-L}{\epsilon}\right\rceil\right)$ iterations.

Finally, we can obtain the overall asypmtotic complexity of Algorithm 3 as $\mathcal{O}(NKI_DI_BI_L)$. Under the assumption that $I_D$, $I_B$, $I_L$ are fixed numbers[4] that do not depend on $K$ and $N$, then the complexity of Algorithm 3 is linear in $K$, and above all in the number of subcarriers $N$. On the other hand, other methods which do not exploit learning theory have a complexity which is either polynomial (typically cubic) in both $N$ and $K$, or linear in $K$ but polynomial in $N$.

### B. Asymptotic complexity of Algorithm 4

The barrier method is similar to Algorithm 3, except for the fact that the feasibility of the generic best-response problem needs to be tested, which is accomplished by solving the *convex* feasibility test in (25). Since the test is convex, it can be solved with standard convex tools with polynomial complexity in the number of variables in the test, which is $N$. In particular, an upper-bound to the complexity of any convex problem is known to scale with the fourth power of the number of variables [49]. Keeping this in mind, and observing that the number of better responses before reaching convergence is proportional to $KI_BI_D$, the asymptotic complexity related to testing the feasibility of the better-responses problem is upper-bounded by $O(N^4KI_BI_D)$. So, the overall complexity of the barrier method can be upper-bounded as $\mathcal{O}(NKI_BI_DI_L)+O(N^4KI_BI_D)$, where the first term can be derived by a similar analysis as for the case without QoS. So, enforcing QoS constraints causes a relevant complexity increase.

### C. Asymptotic complexity of Algorithm 5

As anticipated in Section IV-B, the goal of Algorithm 5 is to reduce the complexity due to testing the feasibility of the QoS constraints. To this end, Algorithm 5 considers per-carrier QoS constraints, thus causing a possible performance loss, but enabling to evaluate the feasibility of each better-response problem by simply checking the condition in (34). This has complexity proportional to $N$, since it requires to compute (33) for all $N$ subcarriers. As a result, the asymptotic complexity of Algorithm 5 can be written as $\mathcal{O}(NKI_BI_DI_L)+O(NKI_BI_D)\approx\mathcal{O}(NKI_BI_DI_L)$, wherein the approximation stems from the fact that $I_L$ is typically of the order of a few tens. In any case, the complexity of Algorithm 5 is linear in $N$.

[4] In line with previous works on radio resource allocation, we assume that the number of iterations $I_D$, $I_B$, $I_L$ does not depend on the problem size, i.e. on $K$ and $N$. This approximation is typically made in related literature since the dependency between the iteration number and the problem size is a very implicit one, and it appears prohibitive to model it mathematically.

## VI. Numerical results

In this section we assess the performance of the proposed solutions through numerical simulation. We assume that $N = 4$ subcarriers are available for transmission, and $K = 12$ users are uniformly distributed over a square area of edge $L = 200$ $m$. The subcarrier bandwidth is set to $W = 10.93$ kHz, and the noise power spectral density is $N_0 = -173$ dBm/Hz for all $n \in \mathcal{N}$. Also, we assume that the total static hardware power dissipated by each transmitter is $P_{c,k} = -20$ dBW for all $k \in \mathcal{K}$. Channel gain coefficients $\alpha_{k,n}$ and $\beta_{k,n}$ are generated according to the path-loss model for Jakes fading [50], and the path-loss coefficient is set to $d = 4$. For each $k \in \mathcal{K}$ and $n \in \mathcal{N}$, we assume that $\xi_{k,n} = \xi \cdot \alpha_{k,n}$ where, unless stated otherwise, $\xi = 0.01$ represents a self-interference coefficient which reflects hardware impairments and/or imperfect channel estimation at the receiver. Finally, we assume that the inefficiency of the power amplifier is constant for all users and over all subcarriers, and it is set to $\mu = 1.02$.

The results presented in the following are averaged over 1000 independent simulation runs.

First, we investigate the effect of self-interference on the system achievable performance. Accordingly, in Fig. 1(a) and Fig. 1(b) we show the system GEE and achievable rate versus the maximum transmission power $P_{max}$, computed for the power allocation output by Algorithm 3, for different values of the self-interference coefficient $\xi$. It is shown that both the GEE and the rate increase as the maximum transmission power level $P_{max}$ increases, and both asymptotically converge towards a saturation point while increasing $P_{max}$. This behavior is expected, since the GEE admits a finite maximum, which is obtained with no need of using all the available power transmission. Thus, when $P_{max}$ is large enough to attain such global maximizer, the transmit power is not increased anymore, as this would lead to a decrease of the GEE value. Moreover, both figures show that self-interference might drastically affect the performance of the network. As an example, Fig. 1(a) shows that the GEE of the system when $\xi = 0$ is twice larger than the GEE achieved when $\xi = 0.1$. Accordingly, Fig. 1(a) and Fig. 1(b) show that the performance of the network are better when the effect of the self-interference is small and limited.

In Fig. 2, we show the average number of iterations needed by each nested loop in Algorithm 3 to reach convergence. Specifically, Fig. 2 reports the average number of times the parameter $\lambda_j$ was updated (Dinkelback Algorithm), the average number of better responses per-user to be computed before the better response dynamics converges (Better Response Dynamics Per-User), and the average number of iterations needed by Algorithm 2 to compute each better-response (Learning Mechanism). For any given $P_{max}$, multiplying the values of the three curves shown in Fig. 2 gives the average number of total iterations (i.e. accounting for all three nested loops) required for convergence. It is shown that the number of iterations is in general low and only a few iterations are needed to reach convergence in Algorithm 3. The number of iterations needed by the learning mechanism is higher than the other two loops, but it is worth noting that each iteration of Algorithm 2 requires only a closed-form variable update. Finally, we observe that the number of iterations of the learning mechanism increases with $P_{max}$, since higher $P_{max}$ produce larger feasible sets over which the optimization is performed.

TABLE I
Ratio between the asymptotic complexity of the method from [16] and our proposed method.

| | $-20$dBW | $-10$dBW | 0dBW | 10dBW |
|---|---|---|---|---|
| Complexity Ratio | $3.025*10^3$ | $2.875*10^3$ | $1.053*10^3$ | 767.778 |

Next, Table I shows the ratio between the asymptotic complexity of the approach from [16], which does not use learning methods[5], and that of the proposed method, as derived in Section V. As for the evaluation of the asymptotic complexity of the method from [16], from [49] we have that a general upper-bound for the complexity of the convex problem solution scales with the fourth power of the total number of variables, that is $KN$ in our case. Then, denoting by $I_S$ and $I_D$ the number of approximate fractional problems to solve and the number of Dinkelbach's iterations for each approximate fractional problem, the complexity of the method from [16] can be evaluated as $\mathcal{O}(I_S I_D K^4 N^4)$. Nevertheless, accounting for the fact that the complexity result from [49] is an upper-bound, in Table I we have considered a cubic complexity for the method from [16], namely $\mathcal{O}(I_S I_D K^3 N^3)$, motivated by the fact that a cubic complexity is what is required to solve a KKT system in the case of linear problems, which is a simpler scenario as the one at hand here. Despite this conservative choice, Table I clearly shows that the proposed method has a much lower complexity than the method from [16]. This result is particularly relevant when taken together with the results in Fig. 5, which will show how the proposed method performs very closely to the method from [16].

Next, we focus on the case when a minimum data-rate requirement $R_{min,k}$ is introduced in the GEE maximization problem. Specifically, we assess the performance of the barrier and generalized games methods for GEE maximization under minimum data-rate requirements proposed in Sections IV-A and IV-B, respectively. Furthermore, we compare them with the algorithm proposed in Section III which does not consider any QoS constraint. Unless otherwise stated, we assume that the barrier cost parameter in (22) used to execute the barrier method is $\rho = 1$. Note that the maximum achievable rate for each user is $R_{max,k} = \sum_{n=1}^{N} \log_2(1 + \alpha_{k,n}/\xi_{k,n})$. Since $\alpha_{k,n}/\xi_{k,n} = 1/\xi = 100$, the maximum achievable rate each user can obtain in the ideal condition in which no multi-user interference is present and infinite transmit power is available is $R_{max,k} = N \log_2(1 + 1/\xi) = 26.63$ bit/s/Hz. Also, we assume $R_{min,k} = R_{min}$ for all $k \in \mathcal{K}$.

As already discussed in Sections IV-A and IV-B, GEE maximization subject to QoS constraints might be unfeasible. However, to provide a fair comparison between the scenarios with and without QoS, if an unfeasibility is detected, in our simulations we slightly relax the QoS requirements. In particular, as for the barrier method, we relax the barrier function by assuming $\varrho_k(\boldsymbol{p}) = C$ if $\sum_{n=1}^{N} \log_2(1+\gamma_{k,n}) < R_{min,k}$, where $C$ is a sufficiently large negative real valued number. In order to set the value of $C$, note that the asymptotic achievable rate of the system is equal to $R_{max} = \sum_{k=1}^{K} R_{max,k} \approx 320$ bit/s/Hz. Also, the first term in (6), which represents the actual data-rate of the system, is always upper-bounded by $R_{max}$, which implies that, to generate a strong penalty term in (24), the absolute value of $C$ should be larger than $R_{max}$. Accordingly, in our simulations we assume that $|C| = 10^3 >> R_{max}$. As for the generalized games method, when an unfeasibility is detected, as discussed in Section IV-B, we enforce the minimum power constraint (33) only on a subset $\mathcal{N}_k^F \subseteq \mathcal{N}$ of subcarriers such that $\sum_{n\in\mathcal{N}_k^F} P_{min,k}^{(n)} \leq P_{max,k}$.

[5] This method merges Dinkelbach's algorithm and sequential optimization, solving a sequence of approximate fractional problems, refining the approximation after each iteration.

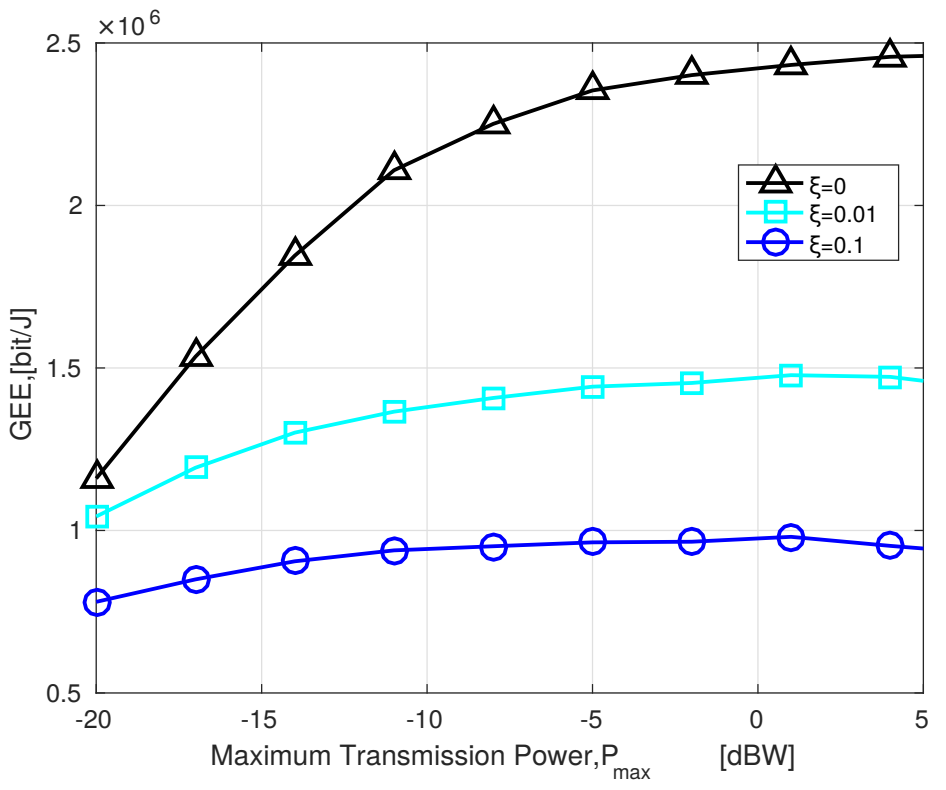

(a) GEE

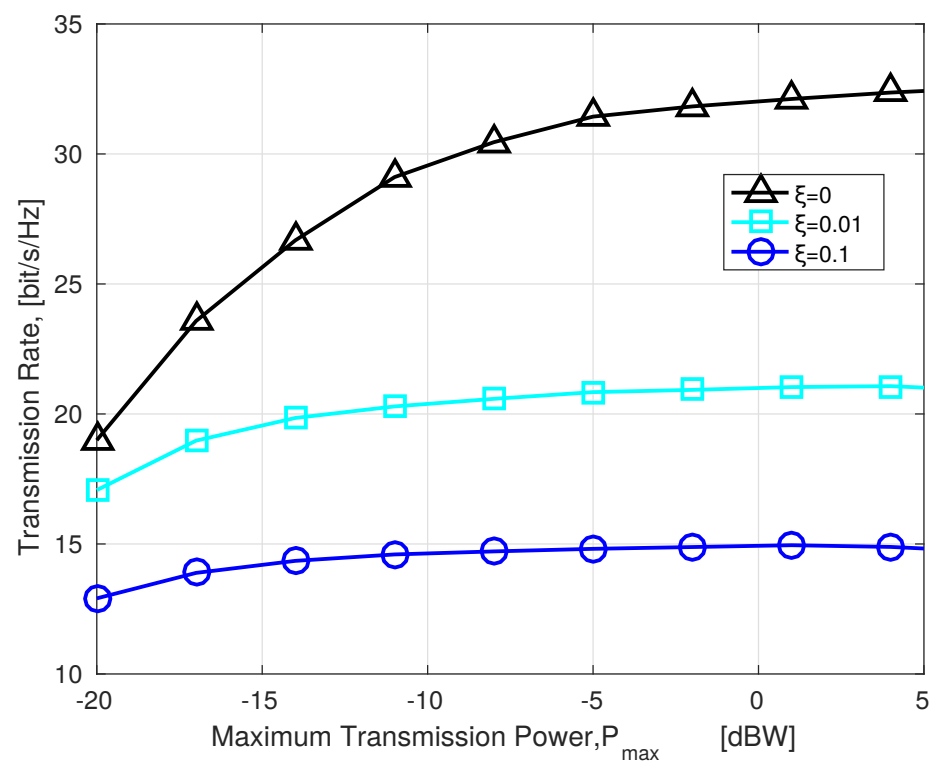

(b) Achievable Rate

Fig. 1. GEE and achievable rate as a function of the maximum transmission power $P_{max}$ for different values of the self-interference coefficient $\xi$.

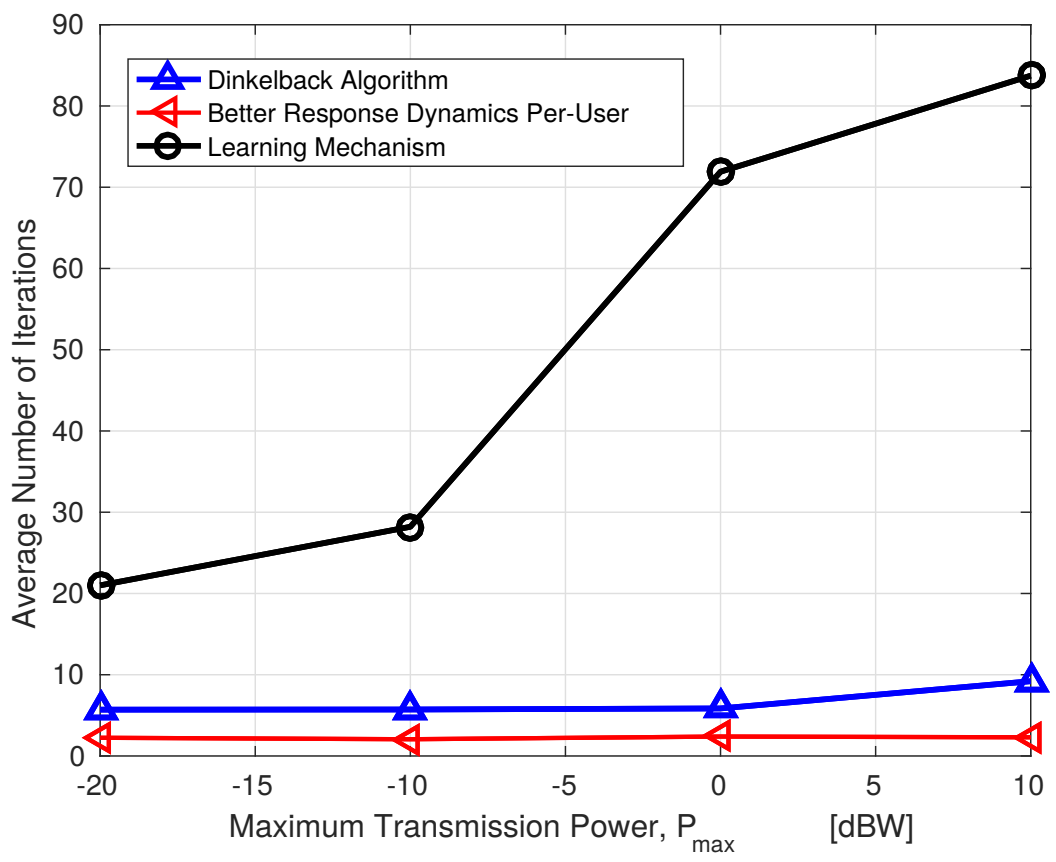

Fig. 2. Average number of iterations needed by each nested loop in Algorithm 3 to reach convergence.

In Fig. 3, we present a comparison between the different proposed algorithms for GEE maximization under minimum QoS requirements as a function of the maximum transmission power $P_{max}$ for different values of the minimum data-rate requirement $R_{min}$. It is worth noting that the algorithm to solve the GEE maximization problem proposed in Section III has not been designed to satisfy any QoS constraints. As a result, Fig. 3(a) shows that the percentage of satisfied users for this algorithm (triangle markers) is low and decreases as the value of $P_{max}$ increases. On the contrary, Fig. 3(a) also shows that the generalized games (circle markers) and barrier (square markers) methods guarantee a higher percentage of satisfied users. Observe that these two methods do not fully meet all the users' QoS requirements, since, as discussed above, the QoS requirements have been slightly relaxed if a feasible solution does not exist.

It is worth noting that the barrier method outperforms all the other algorithms in any of the considered cases as it satisfies a higher number of users, and the percentage of satisfied users with the generalized games method is not considerably impacted by the maximum transmission power level. The GEE and the achievable rate of the system under the three proposed methods are shown in Fig. 3(b) and Fig. 3(c), respectively. The highest value of the GEE is achieved when the algorithm without QoS requirements is considered, since enforcing QoS restricts the problem feasible set. As already shown in Fig. 1(a) and Fig. 1(b), when the value of $P_{max}$ increases, the GEE and the data-rate in this case increase as well. Instead, the GEE in the case of the generalized games method decreases when larger maximum transmission power levels are allowed, while the achievable data-rate remains almost constant. A similar behavior is exhibited by the barrier method, which however results in better performance if compared to the generalized games method. Interestingly, the average per-user achievable rate is almost constant for the two algorithms proposed in Section IV. In conclusion, the obtained results show that the introduction of minimum data-rate requirements decreases the performance of the system in terms of GEE. Actually, in order to provide a minimum QoS level, even those users which experience poor channel conditions have to be scheduled, which inevitably leads to performance degradation in terms of global network energy efficiency.

In Fig. 4, we assess the performance of the barrier method as a function of the barrier cost parameter $\rho$ for different values of the minimum transmission rate $R_{min}$ when $P_{max} = -20$ dBW. From (22), we have that the cost introduced by the barrier method increases when large values of $\rho$ are considered. Accordingly, Fig. 4(a) shows that the percentage of satisfied users increases with the value of $\rho$. As expected, the percentage of satisfied users is higher when the minimum required achievable rate is low. Instead, Fig. 4(b) and Fig. 4(c) show that the GEE and the achievable rate decrease as the cost parameter $\rho$ increases. This stems from the fact that larger values of $\rho$ lead to higher barrier costs, which encourages the system to fulfill the users' rate requirements, at the expense of the system global performance.

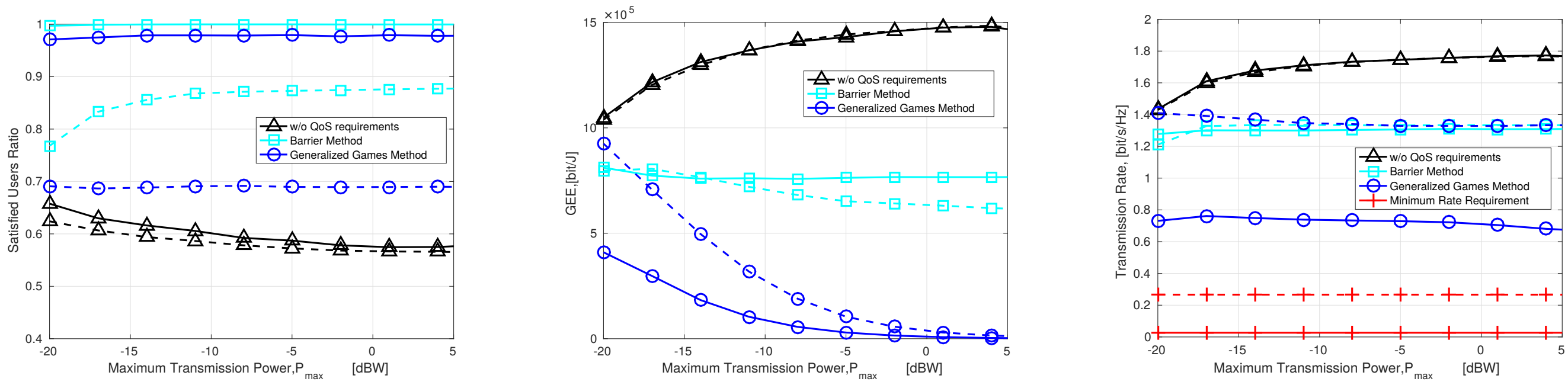

(a) Satisfied Users Ratio (b) GEE (c) Average Per-user Achievable Rate

Fig. 3. Satisfied users ratio, GEE, and average per-user achievable rate as a function of the maximum transmission power $P_{max}$ for different values of the minimum rate $R_{min}$ (Solid lines: $R_{min} = 0.026$ bit/s/Hz; Dashed lines: $R_{min} = 0.266$ bit/s/Hz).

Finally, to assess the performance of the proposed low-complexity solution in Algorithm 3, we compare it with the more sophisticated solution proposed in [16], which enjoys strong optimality properties and has been shown to attain global optimality is several practical scenarios. Let $\text{GEE}^D$ and $\text{GEE}^C$ be the GEE obtained by Algorithm 3 and the approach from [16], respectively, and let the GEE efficiency be defined as the ratio $\text{GEE}^D/\text{GEE}^C$. The closer to one the value of the ratio, the better the performance of the obtained solution. In Fig. 5, we show both the GEE value achieved by the two approaches and the GEE efficiency as a function of $P_{max}$ for different values of the self-interference parameter $\xi$, and with $K = 20$ active users in the system. Remarkably, it is seen that the proposed method performs very close to the approach from [16], despite being less complex. In particular, Algorithm 3 provides the same performance as the method from [16] in almost all considered cases, with the GEE efficiency being lower than one only for small values of the maximum transmission power level.

## VII. Conclusions

This paper has introduced an algorithmic framework for energy-efficient radio resource allocation in multi-carrier wireless interference networks. The maximization of the system GEE subject to both maximum power and minimum rate constraints has been tackled by merging fractional programming, game theory, and learning tools. The resulting framework is provably convergent and strikes a better optimality-complexity trade-off than available alternatives. The merits of the developed framework have been assessed by an extensive numerical analysis, showing that enforcing minimum rate constraints can lead to quite lower GEE values, although enabling supporting minimum communication rates for all users.

## VIII. Appendix

### A. Proof of Proposition 2

*Proof:* To begin with, we observe that many terms in (12) do not depend on $\boldsymbol{p}_{-k}$, and thus are inessential as far as Problem (15) is concerned. Neglecting these terms, Problem (15) can be equivalently restated as follows:

$$\max_{\boldsymbol{p}_k \in \mathcal{S}_k} \sum_{n=1}^{N} \Bigg[ \log_2(1+\gamma_{k,n}) - \nu_{k,n} p_{k,n} + \sum_{i \neq k} \log_2 \Bigg( \sigma_n^2 + \eta_{i,n} p_{i,n} + \beta_{k,n}^{(i)} p_{k,n} + \sum_{\ell \neq i, \ell \neq k} \beta_{\ell,n}^{(i)} p_{\ell,n} \Bigg) \Bigg] \tag{37}$$

with $\nu_{k,n} = \lambda_j \mu_{k,n} - \phi_{k,n}$. Next, we show the uniqueness of the solution by proving that (37) is a strictly concave problem. To this end, let $\hat{V}_k(\boldsymbol{p}_k, \boldsymbol{p}_{-k}, \bar{\boldsymbol{p}}_k)$ be the objective function in (37), and let $\mathbf{H}_k$ be its Hessian matrix with respect to $\boldsymbol{p}_k$. It is easy to show that $\mathbf{H}_k$ is a $N \times N$ diagonal matrix whose $n$-th diagonal element $H_k^{n,n}$ is defined as

$$H_k^{n,n} = \frac{1}{\log(2)} \Bigg[ \frac{\xi_{k,n}^2}{\left(\sigma_n^2 + \xi_{k,n} p_{k,n} + \sum_{i \neq k} \beta_{i,n}^{(k)} p_{i,n}\right)^2} - \frac{\eta_{k,n}^2}{\left(\sigma_n^2 + \eta_{k,n} p_{k,n} + \sum_{i \neq k} \beta_{i,n}^{(k)} p_{i,n}\right)^2} - \sum_{i \neq k} \frac{\beta_{k,n}^{(i)\,2}}{\left(\sigma_n^2 + \eta_{i,n} p_{i,n} + \beta_{k,n}^{(i)} p_{k,n} + \sum_{l \neq i, l \neq k} \beta_{l,n}^{(i)} p_{l,n}\right)^2} \Bigg] \tag{38}$$

Recalling (13), it follows that the difference of the first two terms in (38) is strictly negative if $\alpha_{k,n} > 0$. Similarly, the third term in (38) is strictly negative if $\sum_{i \neq k} \beta_{k,n}^{(i)} \geq 0$, i.e. if $\beta_{k,n}^{(i)} > 0$ for at least one $i$. Hence, unless $\alpha_{k,n} + \sum_{i \neq k} \beta_{k,n}^{(i)} =$ 0, $H_k^{n,n}$ is always negative[6]. It follows that the eigenvalues are negative, and $\mathbf{H}_k$ is definite negative. Thus $\hat{V}_k(\boldsymbol{p}_k, \boldsymbol{p}_{-k}, \bar{\boldsymbol{p}}_k)$ is strictly concave in $\boldsymbol{p}_k$, and it admits a unique (global) maximizer, i.e., Problem (37) has a unique solution.

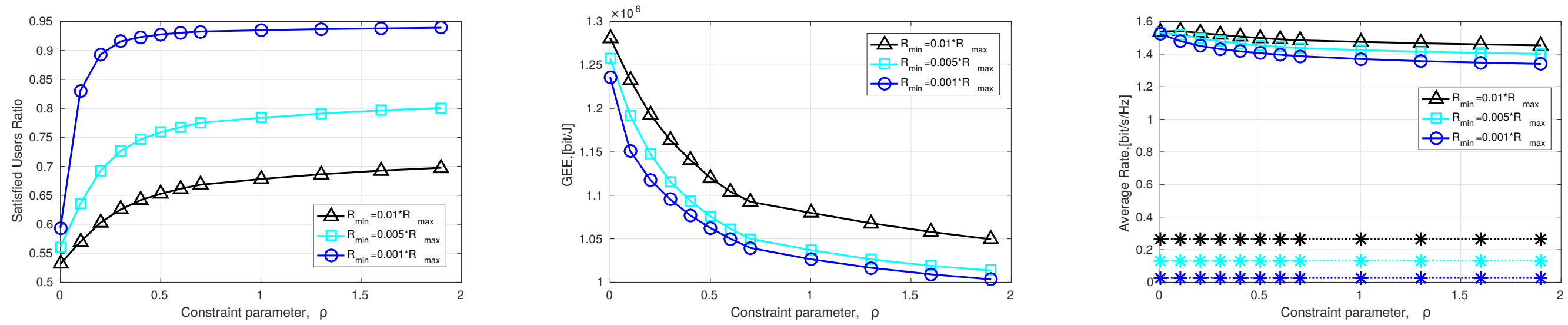

(a) Satisfied Users Ratio (b) GEE (c) Average Per-user Achievable Rate

Fig. 4. Satisfied users ratio, GEE, and average per-user achievable rate as a function of the barrier cost parameter $\rho$ for different values of the minimum rate $R_{min}$.

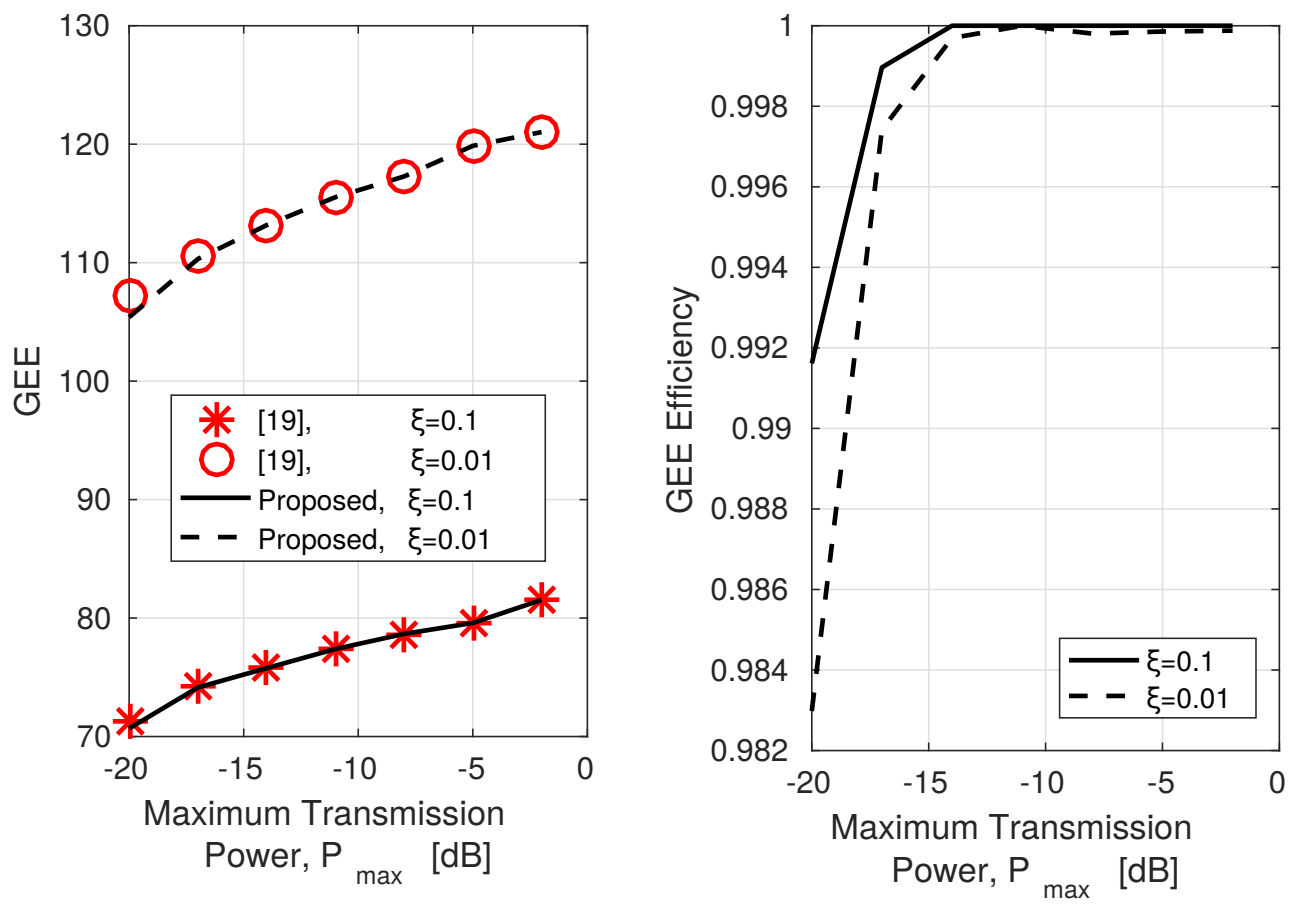

Fig. 5. GEE efficiency as a function of the maximum transmission power $P_{max}$ for different values of the self-interference $\xi$.

Finally, let us show that the unique solution of (37) is a better response for player $k$. Let $\boldsymbol{\pi}_k = \{\pi_{k,n}\}_n$ be a strategy profile for user $k$. By definition, the approximated utility function $\hat{u}_k(\boldsymbol{\pi}_k, \boldsymbol{p}_{-k}, \boldsymbol{\pi}_k)$ in (12) represents the value of the function $u_k(\boldsymbol{\pi}_k, \boldsymbol{p}_{-k})$ in (8) being evaluated at $\boldsymbol{\pi}_k$. Therefore, we have that $u_k(\boldsymbol{\pi}_k, \boldsymbol{p}_{-k}) = \hat{u}_k(\boldsymbol{\pi}_k, \boldsymbol{p}_{-k}, \boldsymbol{\pi}_k)$. Let $\boldsymbol{p}_k$ be the solution of Problem (37) when maximizing $\hat{V}_k(\boldsymbol{p}_k, \boldsymbol{p}_{-k}, \boldsymbol{\pi}_k)$. We have that $\hat{u}_k(\boldsymbol{\pi}_k, \boldsymbol{p}_{-k}, \boldsymbol{\pi}_k) \leq \hat{u}_k(\boldsymbol{p}_k, \boldsymbol{p}_{-k}, \boldsymbol{\pi}_k)$, and thus $u_k(\boldsymbol{\pi}_k, \boldsymbol{p}_{-k}) \leq \hat{u}_k(\boldsymbol{p}_k, \boldsymbol{p}_{-k}, \boldsymbol{\pi}_k)$. Also, we have that $\hat{u}_k(\boldsymbol{p}_k, \boldsymbol{p}_{-k}, \boldsymbol{\pi}_k) \leq u_k(\boldsymbol{p}_k, \boldsymbol{p}_{-k})$ by construction. Thus, it follows that $u_k(\boldsymbol{\pi}_k, \boldsymbol{p}_{-k}) \leq u_k(\boldsymbol{p}_k, \boldsymbol{p}_{-k})$, which implies that $\boldsymbol{p}_k$ is a better-response to $\boldsymbol{p}_{-k}$ for user $k$, and concludes the proof. ■

[6]It is worth noting that $\alpha_{k,n} + \sum_{i \neq k} \beta_{k,n}^{(i)} = 0$ implies that no useful signal can be transmitted by user $k$ on channel $n$, and $p_{n,k} = 0$ is the only feasible strategy for $k$ on these channels. Therefore, these strategies are constant, i.e., they can be removed from the strategy space, and the problem still remains strictly concave in the remaining strategies.

### *B. Proof of Proposition 3*

*Proof:* We first derive the corresponding ordinary differential equation (ODE) for (16) as:

$$\begin{cases} \dot{y}_{k,n} = \hat{v}_{k,n}(p_{k,n}, \boldsymbol{p}_{-k}, \bar{\boldsymbol{p}}_k) \\ p_{k,n} = P_{max,k} \frac{e^{y_{k,n}}}{1 + \sum_{m=1}^{N} e^{y_{k,m}}} \end{cases} \tag{39}$$

where $\dot{y}_{k,n}$ is the first-order time derivative of $y_{k,n}$, and $\hat{v}_{k,n}(p_{k,n})$ is defined in (17).

Since $\hat{v}_{k,n}$ is bounded by definition, it can be easily shown that it is also Lipschitz. Furthermore, from Proposition 2, $\hat{u}_k$ is strictly concave and admits a unique maximizer $\boldsymbol{p}_k^* = \{p_{n,k}^*\}$. For the sake of readability, here we denote the maximum transmission power $P_{max,k}$ of user $k$ as $P_k$.

Let us consider the function $L(\boldsymbol{p}_k)$

$$L(\boldsymbol{p}_k) = P_k \sum_{n=1}^{N} \log\left(\frac{P_k - p_{k,n}^*}{P_k - p_{k,n}}\right) + p_{k,n}^* \log\left(\frac{p_{k,n}^*}{p_{k,n}} \cdot \frac{P_k - p_{k,n}}{P_k - p_{k,n}^*}\right) \tag{40}$$

The first-order time derivative of $L(\boldsymbol{p}_k)$ is $\dot{L} = \frac{dL(\boldsymbol{p}_k)}{dt} = \sum_{n=1}^{N} \hat{v}_{k,n}(p_{k,n})(p_{k,n} - p_{k,n}^*)$, and the strict concavity of $\hat{u}_k$ implies $\dot{L} < 0$. Furthermore, it can be easily shown that $L(\boldsymbol{p}_k^*) = 0$, and $L(\boldsymbol{p}_k) > 0$ for all $\boldsymbol{p}_k \neq \boldsymbol{p}_k^*$. Therefore $L(\boldsymbol{p}_k)$ is a strict Lyapunov function for the ODE in (39). Also, $L(\boldsymbol{p}_k)$ is radially unbounded, which implies that the unique equilibrium of the system, i.e., the unique maximizer $\boldsymbol{p}_k^*$, is also globally asymptotically stable. Let $\boldsymbol{p}_k(t)$ be a solution orbit of (39). The above results guarantee that the ODE in (39) converges towards $\boldsymbol{p}_k^*$ with probability 1, i.e., $\boldsymbol{p}_k(t) \to \boldsymbol{p}_k^*$ as $t \to +\infty$. By decoupling (39) w.r.t. $p_{k,n}$ and $y_{k,n}$, we obtain

$$\dot{p}_{k,n} = \frac{dp_{k,n}}{dt} = p_{k,n}\left(1 - \frac{p_{k,n}}{P_k}\right)\hat{v}_{k,n}(p_{k,n}) \tag{41}$$

A second-order Taylor expansion of (39) gives us

$$p_{k,n}(t+1) = p_{k,n}(t) + \frac{1}{2}\Gamma\gamma_t^2 + \gamma_t p_{k,n}(t)\left(1 - \frac{p_{k,n}(t)}{P_k}\right)\hat{v}_{k,n}(p_{k,n}(t)) \tag{42}$$

where $\Gamma$ is bounded due to the strict concavity of $\hat{u}_k$. Observe that (42) can be seen as a discretized version of (41), up to a bounded error. Since $\sum_t \gamma_t^2 < \sum_t \gamma_t = +\infty$ by assumption, $\boldsymbol{p}_k(t)$ is an asymptotic pseudo-trajectory for (39) [51].

Next, we prove that $p_{k,n}(t)$ converges towards $p_{k,n}^*$ for all $n \in \mathcal{N}$. By rewriting (40) in terms of $\boldsymbol{y}_k = \{y_{k,n}\}$, and by performing a second-order Taylor expansion of $L(\boldsymbol{y}_k(t+1))$, we obtain

$$L(\boldsymbol{y}_k(t+1)) = L(\boldsymbol{y}_k(t)) + \gamma_t \sum_{n=1}^{N} \hat{v}_{k,n}(p_{k,n}(t))(p_{k,n}(t) - p_{k,n}^*) + \frac{1}{2}\Gamma' \gamma_t^2 \tag{43}$$

for some bounded $\Gamma' > 0$. Recall that $\sum_{n=1}^{N} \hat{v}_{k,n}(p_{k,n}(t))(p_{k,n}(t) - p_{k,n}^*) < 0$ by strict concavity, and that $\boldsymbol{p}_k^*$ is globally asymptotically stable. Therefore, there exists a compact set $\mathcal{L}$ which is also a basin of attraction for $\boldsymbol{p}_k^* \in \mathcal{L}$. Then, assume ad absurdum that $p_{k,n}(t)$ does not converge towards $p_{k,n}^*$. If $\boldsymbol{p}_k(t)$ does not converge, it must stay at a bounded distance from $p_{k,n}^*$. That is, there must exist $a > 0$ such that $\sum_{n=1}^{N} \hat{v}_{k,n}(p_{k,n}(t))(p_{k,n}(t) - p_{k,n}^*) \le -a$ for all $t$. Accordingly, (43) can be approximated as $L(\boldsymbol{y}_k(t+1)) \le L(\boldsymbol{y}_k(t)) - \gamma_t a + \frac{1}{2}\Gamma' \gamma_t^2$, which, by telescoping, leads to $L(\boldsymbol{y}_k(t+1)) \le L(\boldsymbol{y}_k(0)) - a\sum_t \gamma_t + \frac{1}{2}\Gamma' \sum_t \gamma_t^2$. Thus, since by assumption $\sum_t \gamma_t^2 < \sum_t \gamma_t = +\infty$, we have that $L(\boldsymbol{y}_k(t+1)) \le -\infty$, which is a contradiction as $L(\boldsymbol{y}_k)$ is lower bounded by construction. Therefore, there must exist a large enough $t'$ such that $\boldsymbol{p}_k(t') \in \mathcal{L}$ and $\lim_{t \to +\infty} \boldsymbol{p}_k(t) = \boldsymbol{p}_k^*$ [51], which concludes the proof. ∎

### C. Proof of Proposition 4

*Proof:* The first step of the proof is to show that whenever $\bar{F}(\lambda_j) \ge 0$, it holds $\lambda_{j+1} \ge \lambda_j$. To see this, denote by $\bar{\boldsymbol{p}}_j$ the equilibrium of the better-response dynamics played at iteration $j$ of Algorithm 3, and define $f_j(\boldsymbol{p}) = \sum_{k=1}^{K}\sum_{n=1}^{N} \log_2(1 + \gamma_{k,n}(\boldsymbol{p}))$, and $g_j(\boldsymbol{p}) = P_c + \sum_{k=1}^{K}\sum_{n=1}^{N} \mu_{k,n} p_{k,n}$. Then, recalling that $g_j(\boldsymbol{p}) \ge 0$ for all $\boldsymbol{p} \in \mathcal{P}$, it holds:

$$0 \le \bar{F}(\lambda_j) = f(\bar{\boldsymbol{p}}_j) - \lambda_j g(\bar{\boldsymbol{p}}_j) \tag{44}$$

$$\Leftrightarrow \frac{\bar{F}(\lambda_j)}{g(\bar{\boldsymbol{p}}_j)} = \frac{f(\bar{\boldsymbol{p}}_j)}{g(\bar{\boldsymbol{p}}_j)} - \lambda_j = \lambda_{j+1} - \lambda_j \ge 0\ . \tag{45}$$

Hence, the sequence $\{\lambda_j\}_j$ is monotonically increasing as long as $\bar{F}(\lambda_j) > 0$. Moreover, $\{\lambda_j\}_j$ is upper-bounded, since $\lambda_j$ is the GEE value (3a) achieved after the $j$-th iteration. Thus, if $\bar{F}(\lambda_j) > 0$ holds until $\{\lambda_j\}_j$ converges, we are in the first case of Proposition 4. Also, as the sequence $\{\lambda_j\}_j$ increases, the sequence $\{F(\lambda_j)\}_j$ decreases. Therefore, it holds:

$$F(\lambda_{j+1}) = \max_{\boldsymbol{p}\in\mathcal{P}}\{f(\boldsymbol{p}) - \lambda_{j+1} g(\boldsymbol{p})\} \le \max_{\boldsymbol{p}\in\mathcal{P}}\{f(\boldsymbol{p}) - \lambda_j g(\boldsymbol{p})\} = F(\lambda_j)\ , \tag{46}$$

Upon convergence, if $F(\lambda) = 0$, Algorithm 3 has achieved global optimality; otherwise, if $F(\lambda) = c$, a suboptimal solution has been attained.

If instead we are in the second case of Proposition 4, the algorithm clearly terminates at iteration $\bar{j}$, since $\bar{F}(\lambda_j) < 0 < \epsilon$. Moreover, by the same steps used to show the first part, it follows that $\{\lambda_j\}_j$ is monotonically increasing for all $0 \le j \le \bar{j}$, whereas $\lambda_{\bar{j}} \ge \lambda_{\bar{j}+1}$. ∎

### D. Proof of Proposition 5

*Proof:* From (35), it is straightforward to verify that $\sum_{n=1}^{N} p'_{k,n}(t+1) \le \Delta_k$. Also, from (38) and Proposition 2, we have that $H_k^{n,n}(p_{k,n}(t+1)) = H_k^{n,n}(p'_{k,n}(t+1) + P_{min,k}^{(n)}) \le 0$. That is, $\hat{u}_k(\boldsymbol{p}'_k(t), \boldsymbol{p}_{-k}, \bar{\boldsymbol{p}}_k)$ is strictly concave w.r.t. $\boldsymbol{p}'_k(t) = \{p'_{k,n}(t)\}_n$ over the shrunken feasible set $\prod_{n=1}^{N}[0, \Delta_k]$, and Problem (32) admits a unique solution w.r.t. $\boldsymbol{p}'_k(t)$. The convergence of (35) towards the solution of Problem (32) can be proved by following the same steps used in the proof of Proposition 3, where $P_{max,k}$ is replaced by $\Delta_k$. ∎

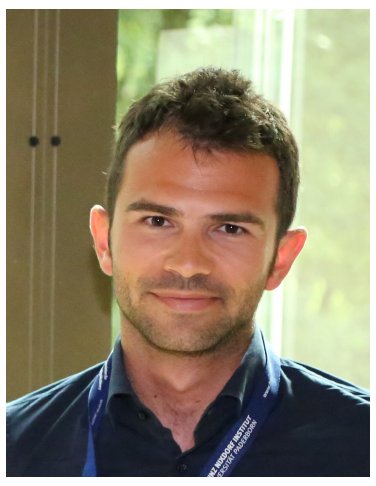

**Salvatore D'Oro** (M'16) received received his Ph.D. degree from the University of Catania in 2015. He is currently a Postdoctoral Researcher at Northeastern University, Boston, USA. In 2015 and 2016, he organized the 1st and 2nd Workshops on COmpetitive and COoperative Approaches for 5G networks (COCOA), and served on the Technical Program Committee (TPC) of Med-Hoc-Net 2018 and the CoCoNet8 workshop at IEEE ICC 2016. In 2013, he served on the TPC of the 20th European Wireless Conference (EW2014). In 2013 and 2015, he was a Visiting Researcher at Universit Paris-Sud 11, Paris, France and at Ohio State University, Ohio, USA. In 2017, he was a Visiting Researcher at Northeastern University. Dr. D'Oro's research interests include game-theory, optimization, learning and their applications to telecommunication networks.

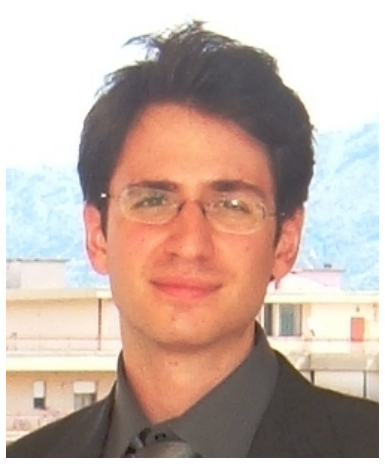

**Alessio Zappone** (SM'16) is a research associate at CentraleSupelec (Gif-sur-Yvette, France). He received his Ph.D. from the University of Cassino and Southern Lazio in 2011 and worked with the Consorzio Nazionale Interuniversitario per le Telecomunicazioni (CNIT) in the framework of the FP7 EU-funded project TREND. From 2012 to 2016, he was the Principal Investigator of the CEMRIN project on energy-efficient resource allocation in wireless networks, funded by the German research foundation (DFG). Since 2016, he has been an adjoint professor at the University of Cassino and Southern Lazio, Cassino, Italy. In 2017, he was the recipient of the H2020 MSCA IF BESMART fellowship. His research interests lie in the area of communication theory and signal processing, with main focus on optimization techniques for resource allocation and energy efficiency maximization. He was the recipient of an FP7-Newcom# grant in 2014 and in 2017 he was selected as an exemplary reviewer for the IEEE TRANSACTIONS ON COMMUNICATIONS and IEEE Transactions on WIRELESS COMMUNICATIONS. He currently serves as an Associate Editor for the IEEE SIGNAL PROCESSING LETTERS and has served as a Guest Editor for the IEEE JOURNAL ON SELECTED AREAS ON COMMUNICATIONS (Special Issue on Energy-Efficient Techniques for 5G Wireless Communication Systems).

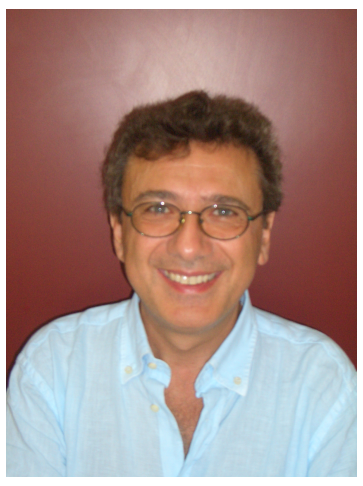

**Sergio Palazzo** received the degree in electrical engineering from the University of Catania, Catania, Italy, in 1977. Since 1987, he has been with the University of Catania, where is now a Professor of Telecommunications Networks. In 1994, he spent the summer at the International Computer Science Institute (ICSI), Berkeley, as a Senior Visitor. In 2003, he was at the University of Canterbury, Christchurch, New Zealand, as a recipient of the Visiting Erskine Fellowship. His current research interests are in modelling, optimization, and control of wireless networks, with applications to cognitive and cooperative networking, SDN, and sensor networks. Prof. Palazzo has been serving on the Technical Program Committee of INFOCOM, the IEEE Conference on Computer Communications, since 1992. He has been the General Chair of some ACM conferences (MobiHoc 2006, MobiOpp 2010), and currently is a member of the MobiHoc Steering Committee. He has also been the TPC Co-Chair of some other conferences, including IFIP Networking 2011, IWCMC 2013, and European Wireless 2014. He also served on the Editorial Board of several journals, including IEEE/ACM Transactions on Networking, IEEE Transactions on Mobile Computing, IEEE Wireless Communications Magazine, Computer Networks, Ad Hoc Networks, and Wireless Communications and Mobile Computing.

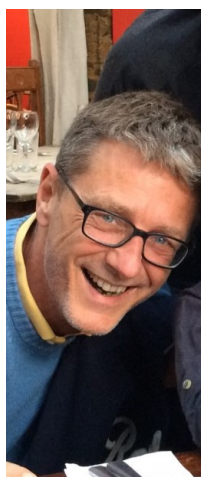

**Marco Lops** was born in Naples, Italy, on March 16, 1961. He received the "Laurea" and the Ph.D. degree from "Federico II" University (Naples), Naples, Italy, where he was an Assistant (1989-1991) and an Associate (1991-2000) Professor. Since March 2000, he has been a Professor at University of Cassino and Southern Lazio and, in 2009-2010, he was full professor at ENSEEIHT (University of Toulouse, France). In fall 2008, he was a Visiting Professor with University of Minnesota and in spring 2009 at Columbia University. He was selected to serve as a Distinguished Lecturer for the Signal processing Society during 2018-2020. His research interests include detection and estimation, with emphasis on communications and radar signal processing.